\documentclass[final,1p,times]{elsarticle}
\usepackage{graphicx}
\usepackage{dcolumn}
\usepackage{bm}
\usepackage{xcolor}
\usepackage{subcaption}
\usepackage{amsmath}
\usepackage{braket}
\usepackage[hyphens]{url}
\usepackage{hyperref} 
\usepackage{cleveref}
\usepackage{multirow}
\usepackage{booktabs}
\usepackage{natbib} 
\setcitestyle{comma,sort&compress,numbers,square}
\hypersetup{hypertex=true,
            colorlinks=true,
            linkcolor=blue,
            anchorcolor=blue,
            citecolor=blue}
\usepackage{amsthm}
\usepackage{geometry}
\usepackage{algpseudocode}
\usepackage{algorithm}
\theoremstyle{definition}

\usepackage{xcolor}
\definecolor{cus-red}{rgb}{0,0,0}
\definecolor{cus-green}{rgb}{0,0,0}
\newcommand{\revone}[1]{{\color{cus-red}#1}}
\newcommand{\revtwo}[1]{\textcolor{black}{{#1}}}
\newcommand{\revthree}[1]{{\color{cus-green}#1}}

\begin{document}
\begin{frontmatter}
\title{Towards Quantum Accelerated Large-scale Topology Optimization}
\date{}
\author[WISC]{Zisheng Ye}
\author[WISC]{Wenxiao Pan\corref{cor}}
\ead{wpan9@wisc.edu}
\cortext[cor]{Corresponding author}
\address[WISC]{Department of Mechanical Engineering, University of Wisconsin-Madison, Madison, WI 53706, USA}

\begin{abstract}
We present an efficient topology optimization (TO) method that not only enhances computational efficiency on classical computing but also provides a practical pathway for leveraging quantum computing to achieve further acceleration. The method targets large-scale, multi-material TO of three-dimensional (3D) continuum structures, beyond prior quantum TO studies limited to small-scale and single-material problems. Building on our discrete-variable TO framework (DVTO-MT), which employs multi-cut optimization and trust regions to reduce iteration counts and thereby PDE solver calls, the proposed method introduces a modified Dantzig–Wolfe (MDW) decomposition to further reduce per-iteration optimization time. The MDW method exploits the block-angular structure of the problem to decompose the mixed-integer linear program (MILP) into reduced-size global and local sub-problems. Evaluations on large-scale 3D bridge design problems demonstrate orders-of-magnitude reductions in computational time, with robust performance even for designs exceeding 50 million variables where classical MILP solvers fail to converge. Furthermore, the computationally intensive local sub-problems are transformed into equivalent quadratic unconstrained binary optimization (QUBO) formulations for quantum acceleration. The resulting QUBOs require only sparse qubit connectivity, a crucial consideration for near-term quantum hardware, and linear construction cost, offering the potential for an additional order-of-magnitude speedup. All observed and estimated speedups become more significant with increasing problem size and when extending from single-material to multi-material designs, highlighting the potential of the proposed method, coupled with quantum computing, to address the scale and complexity of real-world TO challenges.
\end{abstract}


\end{frontmatter}

\section{Introduction}
Spanning diverse sectors, from structural mechanics \cite{wu2021topology}, fluid dynamics \cite{borrvall2003topology, gersborg2005topology, kreissl2011topology}, electromagnetics \cite{kiziltas2003topology}, and photonics \cite{nomura2007structural} to multiphysics applications \cite{dede2009multiphysics, kreissl2010topology, kumar2020topology}, and from  automotive \cite{cavazzuti2011high} to aerospace \cite{zhu2016topology} industries, there is a growing need for designing high-performance, energy-efficient, and lightweight components, devices, and structures. Topology optimization (TO) has emerged as a powerful computational tool to address this need. It determines the optimal distribution of material within a given design domain to maximize structural or functional performance while satisfying the governing physics and design constraints. TO can address discrete truss structures, where the design is of finite dimensions and defined by a finite set of parameters, or continuum structures, where designs are represented as distributed functions. This work focuses on the latter, because it poses greater computational and theoretical challenges. In continuum TO, the governing physics, whether structural mechanics, fluid dynamics, or multiphysics, are typically enforced via state equilibrium equations in the form of partial differential equations (PDEs). When subject to design constraints, e.g., constraining total volume or material usage, the resulting formulation becomes a PDE-constrained optimization problem. The expanding application of TO has brought about substantial computational challenges. These arise from the high-dimensional and complex design space, the intensive cost of solving PDEs, and the need to repeatedly solve PDEs throughout the optimization process. Two key factors largely determine the total computing time: the number of iterations required to achieve convergence, which dictates the number of PDE solver calls; and the time needed to solve the optimization subproblem at each iteration. Therefore, advancing computational methodologies and frameworks to improve TO solution efficiency should focus on reducing iteration counts and accelerating per-iteration solution time.

As a PDE-constrained optimization problem, TO requires discretizing the design domain, typically via finite element methods (FEM), with each element representing a potential material site. This results in an optimization problem with a substantial number of binary (0/1) design variables, where 1 indicates material presence and 0 its absence. The nonlinear dependence of continuous state variables on these binary variables formulates TO as large-scale Mixed-Integer Nonlinear Programming (MINLP) problems, constrained by equality and inequality constraints. Being NP-hard in general \cite{belotti2013mixed}, these problems are computationally intensive using traditional methods \cite{andreassen2011efficient, liang2019topology, picelli2021101, munoz2011generalized,huang2021new,liu2024multi,jia2024fenitop}. Our prior work introduced a TO framework \cite{ye2025discrete} that significantly enhances computational efficiency by reducing optimization iterations and, consequently, the frequency of PDE solving. This framework, referred to as DVTO-MT, integrates generalized Benders' decomposition \cite{ye2023quantum} with adaptive trust regions, formulating the master sub-problem (derived from the MINLP problem) as a multi-cut Mixed-Integer Linear Programming (MILP) problem. This approach enables estimation of the original objective function’s upper and lower bounds, accelerating solution convergence for both convex and non-convex problems across single- and multi-material designs. Compared to widely used traditional methods such as Solid Isotropic Material with Penalty (SIMP) \cite{andreassen2011efficient}, Floating Projection (FP) \cite{huang2021new}, and Sequential Approximate Integer Programming (SAIP) \cite{liu2024multi}, this framework significantly reduces the number of optimization iterations and FEM analyses required, while achieving similar objective function values and optimized topologies. To scale this framework to larger TO problems, advancements to the MILP solver are required. Quantum computing, an emerging paradigm, offers potential superiority over classical computing for combinatorial optimization tasks \cite{albash2018adiabatic, ajagekar2022hybrid, guillaume2022deep, kelany2022quantum,heng2022solve,Liu_LayerVQE_2022}. For example, quantum annealing can accelerate the search for global minima in Quadratic Unconstrained Binary Optimization (QUBO) problems by enabling faster escape from local optima \cite{ray1989sherrington, kadowaki1998quantum}. However, directly converting a MILP problem into an equivalent QUBO problem introduces new challenges. In TO, some design constraints often translate into global constraints within the MILP, which involve all design variables. When mapped into the QUBO formulation, these global constraints produce a dense Hamiltonian, i.e., a QUBO with a densely populated coefficient matrix. This not only imposes the need for all-to-all qubit connectivity, which exceeds the capabilities of most near-term quantum hardware \cite{QCNISQ_Preskill2018,kim2023scalable,dwave-documentation}, but also leads to high pre-processing costs. Specifically, constructing the QUBO's dense coefficient matrix scales quadratically with the number of design variables, diminishing the overall efficiency and offsetting any potential quantum speedup.

Therefore, this paper introduces a new method that significantly enhances the efficiency and scalability of solving MILP problems arising from large-scale TO. It also offers a practical pathway to harness quantum advantages by requiring only sparse qubit connectivity and ensuring that the cost of constructing QUBO scales linearly with problem size. This new method centers on a modified Dantzig-Wolfe (MDW) decomposition. Utilizing its block-angular structure, the MILP problem is decomposed into a series of reduced-size sub-problems, which are categorized as global or local, corresponding to global and local constraints, respectively. Unlike the standard Dantzig-Wolfe decomposition \cite{dantzig1960decomposition}, the MDW formulation is designed to effectively control the size of the global sub-problem. The DVTO-MT framework augmented with MDW can reduce computational time by orders of magnitude and maintains low runtimes even in extreme cases where classical MILP solvers fail to converge, such as designs with over 50 million variables. Furthermore, the local sub-problems, which account for the majority of the computational cost, are essentially Binary Integer Programming (BIP) problems and can potentially be accelerated using quantum computing by translating them into equivalent QUBO problems. Importantly, each local sub-problem involves only local constraints, resulting in QUBO formulations with sparse Hamiltonians. This sparsity eliminates the need for all-to-all qubit connectivity, making the approach well-suited to near-term quantum hardware. Moreover, since each local sub-problem is independent, all QUBOs can be solved in parallel within each MDW iteration, enabling substantial gains in computational speed.

This work represents a distinct advancement over previous efforts that applied quantum computing to TO for improving computational efficiency. Unlike earlier studies limited to discrete truss structures \cite{wang2024mapping, honda2024development}, which represent a small fraction of TO applications, this work addresses more challenging continuum structures. It also surpasses previous work, including our own \cite{ye2023quantum} and others \cite{sukulthanasorn2025novel}, which focused on small-scale, single-material designs and formulated QUBOs with dense Hamiltonians, necessitating all-to-all qubit connectivity and incurring quadratically scaled QUBO construction costs with increasing design variables.

The remainder of this paper is organized as follows. Section \ref{sec:prelimiary} introduces the formulation of the DVTO-MT framework for multi-material TO of 3D continuum structures. Section \ref{sec:method} presents the proposed MDW methodology and its implementation for both classical and quantum computational paradigms. In Section \ref{sec:results}, the proposed approach is evaluated and compared with existing methods in terms of solution quality and computational efficiency through large-scale 3D bridge design, including both single- and multi-material cases across various discretization resolutions. Finally, Section \ref{sec:conclusion} concludes the paper by summarizing the study and highlighting its key contributions.

\section{Preliminary}\label{sec:prelimiary}

\subsection{Formulation of DVTO-MT}\label{subsec:DVTO}

Consider the design of structures within a continuum domain ($\Omega$), for which the design of material distribution is represented by an indicator function $\rho(\mathbf{x})$ with $\mathbf{x}\in \Omega$ \cite{bendsoe2013topology}. The physical equilibrium state of the structures is governed by a partial differential equation (PDE) subject to appropriate boundary conditions (BCs). 
In the context of designing structures governed by linear elasticity, 
the governing PDE is given by:
\begin{equation}
    \begin{aligned}
        \nabla \cdot \bm{\sigma}(\mathbf{u}) + \mathbf{b} =  \mathbf{0} \quad & \text{in} \quad \Omega \;, \\
        \bm{\sigma}(\mathbf{u}) \cdot \mathbf{n} =  \mathbf{h}_\Gamma \quad & \text{on} \quad \Gamma_N \;, \\
        \mathbf{u} =  \mathbf{u}_\Gamma \quad & \text{on} \quad \Gamma_D \;,
    \end{aligned}
    \label{eq:linear_elasticity}
\end{equation}
where $\bm{\sigma}(\mathbf{u}) = \mathbb{C} : \bm{\varepsilon}(\mathbf{u})$ is the stress tensor (2nd order); $\mathbb{C}$ is the fourth-order elasticity tensor; $\bm{\varepsilon} = \frac12(\nabla \mathbf{u} + \nabla \mathbf{u}^\intercal)$ is the strain tensor (2nd order); $\mathbf{b}$ denotes the body force; and $\Gamma = \partial \Omega$ denotes the entire boundary of the design domain with $\Gamma_D$ and $\Gamma_N$ representing the portions where Dirichlet and Neumann BCs are imposed, respectively.  Following numerical discretization of the above PDE (e.g., via FEM), the continuum TO problem in its discrete form can be expressed as:
\begin{equation}
    \begin{aligned}
        \min_{\mathbf{u}, \boldsymbol{\rho}} & \quad f(\mathbf{u}, \bm{\rho}) \\
        \text{s.t.} & \quad \mathbf{K}(\boldsymbol{\rho}) \mathbf{u} = \mathbf{f} \\
        & \quad H_{i_H}(\bm{\rho}) \leq 0, \quad i_H = 1, \dots, n_{H} \\
        & \quad \mathbf{u} \in \mathbb{R}^{3n_v}, \boldsymbol{\rho} \in \{ 0, 1 \}^{n_e\times n_M}
    \end{aligned} 
    \label{eq:non-convex-to}
\end{equation}
where $\mathbf{f}$ encompasses the discretized source terms and BCs; $\mathbf{u}=\{\mathbf{u}_v\}$ with $\mathbf{u}_v =\{u_v, v_v, w_v\}$ denoting the nodal displacement on each vertex of the FEM mesh; $\boldsymbol{\rho}=\{\rho_{e,m}\}$ with $\rho_{e,m}$ denoting the design variable defined at the centroid of each element $e$ in the mesh and for the candidate material $m$; $H_{i_H}$ denotes the inequality constraints that only exerted on design variables $\bm{\rho}$; and $\mathbf{K}$ is the linear operator discretized from Eq. \eqref{eq:linear_elasticity}. The mesh consists of $n_v$ nodal points (or vertices) and $n_e$ elements, with the design subject to $n_M$ candidate materials and $n_H$ inequality constraints.
    
When using FEM to discretize the governing equation in \eqref{eq:linear_elasticity}, the strain tensor $\bm{\varepsilon}_e$ on each element $e$ is calculated from the nodal displacement $\mathbf{u}_v$ as:
\begin{equation}
    \begin{aligned}
    \bm{\varepsilon}_e = &
    \left[
    \begin{matrix}
        \frac{\partial u_v}{\partial x} & \frac{\partial v_v}{\partial y} & \frac{\partial w_v}{\partial z} & \frac{\partial u_v}{\partial y} + \frac{\partial v_v}{\partial x} &    
         \frac{\partial u_v}{\partial z} + \frac{\partial w_v}{\partial x} & \frac{\partial v_v}{\partial z} + \frac{\partial w_v}{\partial y}
    \end{matrix}
    \right]^\intercal 
    =  \begin{bmatrix}
        \frac{\partial}{\partial x} & 0 & 0 & \frac{\partial}{\partial y} & \frac{\partial}{\partial z} & 0 \\
        0 & \frac{\partial}{\partial y} & 0 & \frac{\partial}{\partial x} & 0 & \frac{\partial}{\partial z} \\
        0 & 0 & \frac{\partial}{\partial z} & 0 & \frac{\partial}{\partial x} & \frac{\partial}{\partial y}
    \end{bmatrix}^\intercal
    \begin{bmatrix}
        u_v \\
        v_v \\
        w_v
    \end{bmatrix} = \mathbf{B} \mathbf{u}_v
    \;,
    \end{aligned}
    \label{eq:strain_matrix_3d}
\end{equation}
where $\mathbf{B}$ is the strain-displacement matrix. The elemental stiffness matrix $\mathbf{K}_e$ is then calculated from the elemental elasticity tensor $\mathbb{C}_e$ as:
\begin{equation}
    \mathbf{K}_e = \int_{\Omega_e} \mathbf{B}^\intercal \mathbb{C}_e \mathbf{B} \mathrm{d} \Omega \;,
    \label{eq:elemental_stiffness}
\end{equation}
where
\begin{equation}
    \mathbb{C}_e = \sum_{m=1}^{n_M} \rho_{e, m} \mathbb{C}_m \;,
    \label{eq:elemental_elasticity}
\end{equation}
with the elasticity tensor $\mathbb{C}_m$ in 3D written as: 
\begin{equation}    
    \begin{aligned}       
        &\mathbb{C}_m = 
\frac{E_m}{(1 + \nu_m)(1 - 2\nu_m)}  
     \begin{bmatrix}
        1 - \nu_m & \nu_m & \nu_m & 0 & 0 & 0 \\
        \nu_m & 1 - \nu_m & \nu_m & 0 & 0 & 0 \\
        \nu_m & \nu_m & 1 - \nu_m & 0 & 0 & 0 \\
        0 & 0 & 0 & \frac{1 - 2\nu_m}{2} & 0 & 0 \\
        0 & 0 & 0 & 0 & \frac{1 - 2\nu_m}{2} & 0 \\
        0 & 0 & 0 & 0 & 0 & \frac{1 - 2\nu_m}{2}
    \end{bmatrix} \;.    
    \end{aligned}    \label{eq:element_elasticity_tensor_3d}
\end{equation}
The global stiffness matrix $\mathbf{K}$ in Eq. \eqref{eq:non-convex-to} is finally assembled based on the elemental stiffness matrix $\mathbf{K}_e$ as:
\begin{equation}
    \mathbf{K} = \mathbf{K}_0 + \sum_{e=1}^{n_e} \mathbf{K}_e \;.
    \label{eq:global_stiffness}
\end{equation}
Here, $\mathbf{K}_0$ is augmented to guarantee that the global stiffness matrix $\mathbf{K}$ is symmetric positive definite, noting that for void elements, all entries of $\mathbf{K}_e$ are zero. To minimize the influence of the augmented $\mathbf{K}_0$ on solution accuracy, its value is assigned according to the elasticity tensor $\mathbb{C}_0$ with $E_0 = \varepsilon \max_m(E_m)$ and  $\nu_0 = 0.3$, where $\varepsilon$ is chosen to be a small value, e.g. $10^{-4}$ in the present work. Given the assembled global stiffness matrix $\mathbf{K}$, the nodal displacement $\mathbf{u}$ is obtained by solving the linear system $\mathbf{K} \mathbf{u} = \mathbf{f}$.

In the present work, the inequality constraint $H_{i_H}$ is considered to be the mass constraint commonly employed in multi-material TO design, which limits the total mass of the resulting structure composed of multiple candidate materials, and is generally expressed as:
\begin{equation}
    \begin{aligned}
        & H_{M_0}(\bm{\rho}) = \sum_{e=1}^{n_e} \sum_{m=1}^{n_M} \dfrac{\bar{M}_m}{n_e} \rho_{e, m} \leq \bar{M}_\text{max} \\
        & H_{M_e}(\bm{\rho}) = \sum_{m=1}^{n_M} \rho_{e, m} \leq 1, \quad e = 1, 2, \dots, n_e 
    \end{aligned}
    \label{eq:mass_constraint}
\end{equation}
where $n_M$ candidate materials has a different normalized density $\bar{M}_m$. The first line in the constraint denotes the global constraint for all design variables, ensuring that the total mass fraction of the resulting structure does not exceed $\bar{M}_\text{max}$, the maximum permissible total mass fraction. The second line imposes a local constraint that restricts each element to be occupied by at most one material. Therefore, the mass constraint in Eq. \eqref{eq:mass_constraint} essentially includes $(1+n_e)$ inequality constraints in total. If only a single material is involved, its normalized density is simply $\bar{M}_1 = 1.0$, and hence the mass constraint in Eq. \eqref{eq:mass_constraint} reduces to a volume constraint as:
\begin{equation}
    \sum_{e = 1}^{n_e} \dfrac{\rho_e}{n_e} \leq V_T \;,
    \label{eq:volume_constraint}
\end{equation}
which limits the total volume fraction of the single material used within the design domain $\Omega$ to not exceed the target value $V_T$.

The continuum TO problem in its discrete form as in Eq. \eqref{eq:non-convex-to} is essentially a MINLP problem. In our prior work \cite{ye2025discrete}, we developed the DVTO-MT framework based on a multi-cut formulation and adaptive trust regions for efficiently solving it. This framework serves as the starting point for the methodology proposed in the present work. Under this framework, at each iteration step (e.g., the $k$-th iteration) we solve a primal problem:
\begin{equation}
    \mathbf{u}^k = \mathbf{K}^{-1}(\boldsymbol{\rho}^{k}) \mathbf{f} \;,
    \label{eq:to_fem}
\end{equation}
and a master problem:
\begin{equation}
    \begin{aligned}
        \min_{\boldsymbol{\rho}, \eta} & \quad \eta \\
        \text{s.t.} & \quad \widetilde{f}^j(\bm{\rho}) \leq \eta, \quad \forall j \in \mathcal{P}_s \\
        & \quad t^j(\bm{\rho}) \leq d^j, \quad \forall j \in \mathcal{P}_s \\
        & \quad H_{i_H}(\bm{\rho}) \leq 0, \quad i_H = 1, \dots, n_H \\
        & \quad \boldsymbol{\rho} \in \{0, 1\}^{n_e\times n_M} \;.
    \end{aligned}
    \label{eq:branched_multi-cut}
\end{equation}
The master problem in \eqref{eq:branched_multi-cut} is an optimization problem, which uses trust regions and multiple linear cuts to find the estimated upper bound $\eta$ to the original problem \eqref{eq:non-convex-to}. Each cut $\widetilde{f}^j(\bm{\rho})$, obtained from the sensitivity analysis \cite{andreassen2011efficient}, represents a linear approximation to the original objective function as: 
\begin{equation*}
    \widetilde{f}^j (\bm{\rho}) = f(\mathbf{u}^j, \bm{\rho}^j) + \sum_{e=1}^{n_e} \left( \sum_{m=1}^{n_M} \left[ {\widetilde{w}_{e,m}^j} (\rho_{e, m} - \rho^{j}_{e,m}) \right] \right) \;,
\end{equation*}
where $\widetilde{w}_{e,m}^j$ is the sensitivity of the objective function with respect to the design variable $\rho_{e, m}$ and will be discussed with more details in the next section. The accuracy of such linear approximation is controlled by its corresponding trust region: 
\begin{equation*}
        t^j(\bm{\rho}) =  \frac{1}{n_M n_e} \sum_{e = 1}^{n_e} \left[ \left( 1 - 2 \sum_{m=1}^{n_M} \rho_{e, m}^j \right) \left( \sum_{m=1}^{n_M} \rho_{e, m} \right) + \left( \sum_{m=1}^{n_M} \rho_{e, m}^j \right)^2 \right] \leq d^j \;,       
    \label{eq:move_limit}
\end{equation*}
where $d^j$ is the radius of the trust region.

In Eq. \eqref{eq:branched_multi-cut}, $\mathcal{P}_s$ can include one or multiple indices $j$ with $j \in \{0, 1, \cdots, k-1\}$, which thereby permits leveraging the solutions and sensitivity analysis obtained in prior iteration steps to expedite the solution convergence, as discussed in our prior work \cite{ye2025discrete}. If only one cut is necessary to be included, in this case $\mathcal{P}_s = \{k - 1\}$, the multi-cut formulation in \eqref{eq:branched_multi-cut} can be reduced to the following single-cut formulation:
\begin{equation}
    \begin{aligned}
        \min_{\boldsymbol{\rho}, \eta} & \quad \eta \\
        \text{s.t.} & \quad \widetilde{f}^{k-1}(\bm{\rho}) \leq \eta \\
        & \quad t^{k-1}(\bm{\rho}) \leq d^{k-1} \\
        & \quad H_{i_H} (\bm{\rho}) \leq 0, \quad i_H = 1, \dots, n_H \\
        & \quad \boldsymbol{\rho} \in \{0, 1\}^{n_e\times n_M} \;.
    \end{aligned}
    \label{eq:single-cut}
\end{equation}
A branching procedure \cite{ye2025discrete} automatically determines either \eqref{eq:branched_multi-cut} or \eqref{eq:single-cut} is solved for the master problem at each iteration step. The early optimization stage typically employs $\eqref{eq:single-cut}$, whereas $\eqref{eq:branched_multi-cut}$ is usually invoked during the later stages for finer topology adjustments.

The primal problem \eqref{eq:to_fem} involves solving a large-scale linear system of equations resulting from FEM discretization, refer to as the FEM solution process. At each iteration step $k$, the material configuration $\bm{\rho}^k$ is fixed, and the FEM solution leads to the updated displacement field $\mathbf{u}^k$. The number of times the FEM solver is invoked, denoted as $N_\text{FEM}$, is equal to the total iteration count ($N_\text{Iter}$) required to solve the original problem \eqref{eq:non-convex-to} and also the number of times the master problem \eqref{eq:branched_multi-cut} or \eqref{eq:single-cut} is solved.

The iterative process of solving the primal problem $\eqref{eq:to_fem}$ and the master problem $\eqref{eq:branched_multi-cut}$ or \eqref{eq:single-cut} yields successive upper and lower bounds for the objective function in $\eqref{eq:non-convex-to}$. To terminate the optimization, we minimize the gap between the lowest upper bound $U = \min_{j}(f(\mathbf{u}^j, \bm{\rho}^j))$ for $j\leq k$ and the lower bound $\eta^k$ until it meets the condition $\tfrac{|\eta^k - U|}{|U|} < \xi$, with $\xi$ being the preset tolerance and set to $10^{-2}$ in this work.

\subsection{Sensitivity Analysis and Filtering}
The gradient of the elemental stiffness matrix $\mathbf{K}_e$ with respect to the design variable $\rho_{e, m}$ is calculated as \cite{sun2022sensitivity,ansola2018sequential,SIVAPURAM2018Topology}:
\begin{equation}
    \dfrac{\partial \mathbf{K}_e}{\partial \rho_{e, m}} = \int_{\Omega_e} \mathbf{B}^\intercal \dfrac{\partial \mathbb{C}_e}{\partial \rho_{e, m}} \mathbf{B} \mathrm{d} \Omega = \int_{\Omega_e} \mathbf{B}^\intercal \mathbb{C}_m \mathbf{B} \mathrm{d} \Omega \;.
    \label{eq:stiffness_gradient_elemental}
\end{equation}
To evaluate the sensitivity of the objective function with respect to the design variable $\rho_{e, m}$, the Lagrangian of the optimization problem is first defined as:
\begin{equation}
    \mathbb{L}(\mathbf{u}, \bm{\rho}, \bm{\mu}) = f(\mathbf{u}, \bm{\rho}) + \bm{\mu}^\intercal [\mathbf{K}(\bm{\rho}) \mathbf{u} - \mathbf{f}] \;,
    \label{eq:lagrangian}
\end{equation}
where $\bm{\mu} \in \mathbb{R}^{3n_v}$ is the adjoint variable. According to the Karush–Kuhn–Tucker(KKT) conditions \cite{nocedal1999numerical}, the adjoint variable $\bm{\mu}$ takes the form:
\begin{equation}
    \bm{\mu} = - \mathbf{K}^{-1} \dfrac{\partial f}{\partial \mathbf{u}} \;,
\end{equation}
with which, the gradient of the objective function with respect to the design variable $\rho_{e, m}$ is calculated as:
\begin{equation}
    \begin{aligned}
        \dfrac{\partial f}{\partial \rho_{e, m}} = &  \left(\dfrac{\partial f}{\partial \mathbf{u}}\right)^\intercal \cdot \dfrac{\partial \mathbf{u}}{\partial \rho_{e, m}} 
        = \left(\dfrac{\partial f}{\partial \mathbf{u}}\right)^\intercal \dfrac{\partial \mathbf{K}^{-1}}{\partial \rho_{e, m}} \mathbf{f} 
        =  - \left(\dfrac{\partial f}{\partial \mathbf{u}}\right)^\intercal (\mathbf{K}^{-1})^\intercal \dfrac{\partial \mathbf{K}}{\partial \rho_{e, m}} \mathbf{K}^{-1} \mathbf{f} 
        =  - \left(\mathbf{K}^{-1} \dfrac{\partial f}{\partial \mathbf{u}}\right)^\intercal \cdot \dfrac{\partial \mathbf{K}}{\partial \rho_{e, m}} \mathbf{K}^{-1} \mathbf{f} \\
        = & \bm{\mu}^\intercal \dfrac{\partial \mathbf{K}}{\partial \rho_{e, m}} \mathbf{u} =  \bm{\mu}^\intercal \dfrac{\partial \mathbf{K}_e}{\partial \rho_{e, m}} \mathbf{u} 
        = \bm{\mu}^\intercal \left(\int_{\Omega_e} \mathbf{B}^\intercal \mathbb{C}_m \mathbf{B} \mathrm{d} \Omega \right) \mathbf{u}
    \end{aligned} \;.
\end{equation}
This relation holds true whether the objective function is convex or non-convex. Finally, the sensitivity of the objective function with respect to $\rho_{e, m}$ can be calculated as:
\begin{equation}
    w_{e, m} =
    \left\{
        \begin{aligned}
            & \dfrac{\partial f}{\partial \rho_{e, m}} & & \text{if } \rho_{e, m} = 1 \\
            & \dfrac{2 \dfrac{\partial f}{\partial \rho_{e, m}} \cdot \dfrac{\partial f}{\partial \rho_{e, m^\prime}}}{\dfrac{\partial f}{\partial \rho_{e, m}} + \dfrac{\partial f}{\partial \rho_{e, m^\prime}}} & & 
            \begin{aligned}
                & \text{if } \rho_{e, m} = 0, \\
                & \rho_{e, m^\prime} = 1, m \neq m^\prime 
            \end{aligned} \\
            & \bm{\mu}^\intercal \left(\int_{\Omega_e} \mathbf{B}^\intercal \mathbb{C}_0 \mathbf{B} \mathrm{d} \Omega \right) \mathbf{u} & & \text{if } \sum_{m=1}^{n_M} \rho_{e, m} = 0
        \end{aligned}
    \right. 
    \;.
    \label{eq:objective_sensitivity}
\end{equation}
Here, while the sensitivity is directly determined as the objective function's derivative for $\rho_{e, m} = 1$, it needs to be regulated for $\rho_{e, m} = 0$ \cite{sun2022sensitivity}. Specifically, two scenarios are considered for regularization: 1) when $\sum_{m=1}^{n_M} \rho_{e, m} = 0$, indicating that element $e$ is a void element, the sensitivity is calculated using the elasticity tensor $\mathbb{C}_0$ discussed in \S\ref{subsec:DVTO}; 2) when $\rho_{e, m} = 0$ and $\rho_{e, m^\prime} = 1$ for $m \neq m^\prime$, indicating element $e$ is occupied by material $m^\prime$ but not $m$, the sensitivity is evaluated by the harmonic mean of the gradients with respect to material $m$ and $m^\prime$, respectively. The choice of harmonic mean rather than the arithmetic mean is inspired by Reuss model \cite{reuss1929berechnung} introduced for composite material design. And our numerical experiments find that the harmonic mean is superior for selecting the optimal material in multi-material design.

Furthermore, to mitigate the so-called ``checkerboard" artifact arising from FEM with uniform meshing \cite{lazarov2011filters,liang2019topology}, 
filtering through the radius filter \cite{picelli2021101,sun2022sensitivity} is employed to smooth the sensitivity of the objective function with respect to $\rho_{e, m}$ as:
\begin{equation}
    \widetilde{w}_{e, m} = \dfrac{\sum_{e^\prime \in \mathcal{N}_e^r} h_{e, e^\prime}(r) w_{e, m}}{\sum_{e^\prime \in \mathcal{N}_e^r} h_{e, e^\prime}(r)}\;,
    \label{eq:sensitivity_filtering}
\end{equation}
where $h_{e, e^\prime}(r) = \max(0, r - \|\mathbf{x}_e - \mathbf{x}_{e^\prime}\|_2)$; $e'$ denotes a neighbor element of $e$; and $r$ is the prescribed radius and set to $3h$ in this work with $h$ the mesh's side length. 

\section{Proposed Methodology}\label{sec:method}
\revtwo{In the original DVTO-MT framework \cite{ye2025discrete}, the master problems \eqref{eq:branched_multi-cut} and \eqref{eq:single-cut} are solved by directly calling a classical MILP solver from Gurobi \cite{gurobi}, which relies on a branch-and-bound approach.} While this approach is feasible for smaller 2D designs, the MILP solver's exponentially increasing cost can render it intractable for large-scale 3D problems, as demonstrated by our numerical tests in Sections \ref{subsec:results-single} and \ref{subsec:results-multi}. 

\subsection{Modified Dantzig-Wolfe (MDW) Decomposition}
\label{sec:dw}
The mass constraint $H_{M_0}(\boldsymbol{\rho})\le\bar{M}_\text{max}$ in Eq. \eqref{eq:mass_constraint} and the cuts $\widetilde{f}^j(\bm{\rho})\le \eta$ and the trust regions $t^j(\bm{\rho})\le d^j$ in Eq. \eqref{eq:branched_multi-cut} involve all design variables, and hence are categorized as \textit{global} constraints. Whereas, the material usage constraint $H_{M_e}(\bm{\rho})\le 1$ in \eqref{eq:mass_constraint} only applies to $n_M$ variables on each element, and hence can be regarded as a \textit{local} constraint. As a result, the programming in Eq. \eqref{eq:branched_multi-cut} or \eqref{eq:single-cut} shows a block-angular structure, and thereby can be decomposed into a series of reduced-size sub-problems through the proposed MDW decomposition. Similar to the standard DW decomposition \cite{dantzig1960decomposition}, the reduced-size sub-problems are categorized as global or local, corresponding to global and local constraints, respectively. However, \revtwo{we cannot simply apply the standard DW decomposition, because it typically collects all extreme points, which are generated from the local sub-problems, into the global sub-problem \cite{dantzig1960decomposition}. The ``extreme points" herein refer to the points satisfying the constraints of the local sub-problems. For the programming in Eq. \eqref{eq:branched_multi-cut} or \eqref{eq:single-cut} associated with the TO problems, the number of extreme points generated from each local sub-problem scales exponentially with respect to the number of design variables involved in the local sub-problem. As the number of design variables increases, the resultant global sub-problem could become intractable.} Therefore, different from the standard DW decomposition, our proposed MDW decomposition generates extreme points iteratively by updating the local sub-problems with the Lagrangian multipliers evaluated from the global sub-problem. Instead of preparing the extreme points from the local sub-problems, we search for the optimal Lagrangian multipliers as the size of Lagrangian multipliers would not scale with the number of design variables. The Lagrangian multipliers can be well initiated from a reduced-size problem (corresponding to employing a lower discretization resolution) given the TO problem is constrained by valid physical constraints, as discussed later in \S\ref{subsubsec:initiation}. Given the topological similarity across different discretization resolutions, only a limited number of elements need to be modified compared to the solution obtained from a lower discretization resolution, suggesting that only a small set of extreme points are required to find the optimal solution. This MDW decomposition effectively controls the size of the global sub-problem, which in turn ensures efficiency when employed to solve the master problem \eqref{eq:branched_multi-cut} or \eqref{eq:single-cut}. Although proposed in the context of TO problems, this MDW decomposition is generally applicable to large-scale binary programming with non-overlapping local constraints whose size would not scale with respect to the number of variables.

The derivation uses the mass constraint in Eq. \eqref{eq:mass_constraint} as the general case, with the volume constraint in Eq. \eqref{eq:volume_constraint} treated as a special case handled similarly. To proceed, all design variables are re-indexed with:
\begin{equation}
    \widetilde{\rho}_e = \sum_{m=1}^{n_M} m \rho_{e, m} \;,
    \label{eq:rho_tilde}
\end{equation}
which assigns a new value $\widetilde{\rho}_e$ for all design variables $\rho_{e, m}$ defined on each element $e$. All $\widetilde{\rho}_e$ are then sorted in a non-descending order, thereby providing each element $e$ a new index $e^\prime$. Parallel execution is possible for the sorting task, necessitating communication of the possible values for $\widetilde{\rho}_e$ among processors. Because $\rho_{e, m}$ is binary, $\widetilde{\rho}_e$ is limited to a finite set of possible values, including $0, 1, 2, \dots, n_M$. This feature leads to substantial savings in communication costs, unlike general sorting algorithms that commonly require communicating sub-vectors of $\bm{\rho}$.
 
Next, with the new index $e^\prime$ assigned to each element, we further evenly partition all design variables into $n_\mathcal{D}$ blocks with each block labeled by:
\begin{eqnarray}
    \mathcal{D}_i = \left\{ e \left| \left\lfloor i \cdot \dfrac{n_e}{n_\mathcal{D}} \right\rfloor \leq e^\prime \leq \left\lfloor (i+1) \cdot \dfrac{n_e}{n_\mathcal{D}} \right\rfloor \right. \right\} \;.
    \label{eq:dw-block}
\end{eqnarray}
Eq. \eqref{eq:dw-block} implies that the $i$-th block collects the indices of the elements $e$ with the same or close value of $\widetilde{\rho}_e$, where $\lfloor \cdot \rfloor$ denoting the floor function. As such, each block manages about $\dfrac{n_e n_M}{n_\mathcal{D}}$ design variables. The vector $\bm{\rho}_i$ is formed such that $\bm{\rho}_i = \left\{ \rho_{e, m} | \forall e \in \mathcal{D}_i, m = 1, \dots n_M \right\}$. \revone{This indexing and partitioning strategy can ensure that all design variables $\rho_{e, m}$ associated with a given element $e$ are grouped into the same block and thus included within the same local sub-problem. This is critical for multi-material design problems, as partitioning the design variables defined on the same element into different blocks would no longer preserve the block-angular structure of the problem which is required by the proposed MDW decomposition method. Alternative partitioning strategies, such as spatially-based partitioning, may also be employed, provided they guarantee that all design variables associated with the same element remain within the same block. For single-material design problems, however, only one design variable is defined per element, so the block-angular structure is inherently preserved regardless of the partitioning strategy.}

Following partitioning the design variables into $n_\mathcal{D}$ blocks, the MDW decomposition yields two types of sub-problems: $n_\mathcal{D}$ local sub-problems (named as they are related to the local constraint) and one global sub-problem (named as it is related to all global constraints). Each local sub-problem's size depends solely on the size of $\mathcal{D}_i$. In contrast, the global sub-problem's size is determined by both the number of blocks ($n_\mathcal{D}$) and the number of iterations ($L$) required by the MDW decomposition, exhibiting a scaling of $O(n_\mathcal{D}^2 L^2)$ which is estimated from $\sum_{l=1}^{L} n_{\mathcal{D}}^2 l$. However, because the number of iterations ($L$) required by the MDW decomposition is independent of the number of design variables, as also evidenced in our numerical experiments, the global sub-problem's size can remain small even for large-scale TO problems.

\subsubsection{Local and Global Sub-problem Formulations}

The programming \eqref{eq:branched_multi-cut} involves a single continuous variable $\eta$. Thus, a series of variables $\eta_i$ with $i = 1, \dots, n_\mathcal{D}$ are introduced to replace $\eta$, and thereby \eqref{eq:branched_multi-cut} is rewritten as:
\begin{equation}
    \begin{aligned}
        \min_{\boldsymbol{\rho}, \eta}  \quad & \sum_{i=1}^{n_\mathcal{D}} \eta_i \\
        \text{s.t.} \quad & \widetilde{f}^j(\bm{\rho}) \leq \sum_{i=1}^{n_\mathcal{D}} \eta_i, \quad & & \forall j \in \mathcal{P}_s \\
        & t^j(\bm{\rho}) \leq d^j, \quad & & \forall j \in \mathcal{P}_s \\
        & H_{M_0}(\bm{\rho}) \leq \bar{M}_{\max} \\
        & \sum_{m=1}^{n_M} \rho_{e, m} \leq 1, \quad & & e = 1, 2, \dots, n_e \\
        & \boldsymbol{\rho} \in \{0, 1\}^{n_e\times n_M} \;.
    \end{aligned}
    \label{eq:multi-cuts-dw-original}
\end{equation}
The MDW decomposition for \eqref{eq:multi-cuts-dw-original} leads to $n_D$ local sub-problems, with the $i$-th one ($i=1, 2, \dots, n_D$) given by:
\begin{equation}
    \begin{aligned}
        \min_{\boldsymbol{\rho}_i, \eta_i} \quad & \eta_i  + \sum_{j \in \mathcal{P}_s} \pi^f_j(\widetilde{f}^j(\bm{\rho}_i) - \eta_i)  + \sum_{j \in \mathcal{P}_s} \pi_j^t t^j(\bm{\rho}_i) + \pi^M H_{M_0}(\bm{\rho}_i) \\
        \text{s.t.} \quad & \sum_{m=1}^{n_M} \rho_{e, m} \leq 1, \quad e \in \mathcal{D}_i \\
        & \boldsymbol{\rho} \in \{0, 1\}^{| \mathcal{D}_i | \times n_M} \;.
    \end{aligned}
    \label{eq:multi-cuts-dw-sub}
\end{equation}
Given that $\eta_i$ is unbounded, $\sum_{j \in \mathcal{P}_s} \pi_j^f$ must be equal to 1 to guarantee a bounded objective function in \eqref{eq:multi-cuts-dw-sub} and in turn a feasible solution to \eqref{eq:multi-cuts-dw-sub}. Thus, Eq. \eqref{eq:multi-cuts-dw-sub} can be rewritten as:
\begin{equation}
    \begin{aligned}
        \min_{\boldsymbol{\rho}} \quad & \sum_{j \in \mathcal{P}_s} \pi^f_j\widetilde{f}^j(\bm{\rho}_i) + \sum_{j \in \mathcal{P}_s} \pi_j^t t^j(\bm{\rho}_i) + \pi^M H_{M_0}(\bm{\rho}_i) \\
        \text{s.t.} \quad & \sum_{m=1}^{n_M} \rho_{e, m} \leq 1, \quad e \in \mathcal{D}_i \\
        & \boldsymbol{\rho} \in \{0, 1\}^{| \mathcal{D}_i | \times n_M} \;,
    \end{aligned}
    \label{eq:multi-cuts-dw-sub-reduced}
\end{equation}
where $\pi^f_j$, $\pi^t_j$, and $\pi^M$ are the Lagrange multipliers obtained from solving the global sub-problem, as explained later. The local sub-problem \eqref{eq:multi-cuts-dw-sub-reduced} is essentially a binary programming problem. While classical computing addresses this problem with Branch and Bound (B\&B)-equipped MILP solvers, its NP-complete nature suggests that quantum computing could offer a more efficient solution, as discussed later in \S\ref{subsec:QI}. Further, $n_{\mathcal{D}}$ such problems can be solved simultaneously in parallel, saving computing time and enabling an efficient parallel implementation. The solution to all local sub-problems \eqref{eq:multi-cuts-dw-sub-reduced} in the $l$-th iteration of the MDW decomposition is denoted as $\rho_{e, m}^{k,l}$.

The global sub-problem resulted from the MDW decomposition for \eqref{eq:branched_multi-cut} can then be written as:
\begin{equation}
    \begin{aligned}
        \min_{\boldsymbol{\lambda}, \eta} \quad & \eta \\
        \text{s.t.} \quad & \widetilde{f}^j(\bar{\bm{\rho}}) \leq \eta, \quad & & \forall j \in \mathcal{P}_s \\
        & t^j(\bar{\bm{\rho}}) \leq d^j, \quad & & \forall j \in \mathcal{P}_s \\
        & H_{M_0}(\bar{\bm{\rho}})\leq \bar{M}_\text{max} \\
        & \sum_{q = 1}^{l} \lambda_{i}^q = 1 & & \quad i = 1, 2, \dots, n_\mathcal{D}\\
        & \bm{\lambda} \in [0, 1]^{n_{\mathcal{D}} \times l} \;,
    \end{aligned}
    \label{eq:multi-cuts-dw-master}
\end{equation}
where $\bar{\bm{\rho}}=\{\bar{\rho}_{e, m}\}$ with $\bar{\rho}_{e, m} = \sum_{q = 1}^l \rho_{e, m}^{k,l} \lambda_{i}^q$, $\forall e \in \mathcal{D}_i, m = 1, \dots n_M$, and $i = 1, 2, \dots, n_\mathcal{D}$. It is a linear programming problem, with both $\bm{\lambda}$ and $\eta$ continuous variables, and can be easily solved by a classical linear programming solver, e.g., Gurobi Optimizer \cite{gurobi} used in this work. During the solution of the global sub-problem \eqref{eq:multi-cuts-dw-master}, the Lagrange multipliers $\pi^f_j$, $\pi^t_j$, and $\pi^M$ are determined concurrently with $\boldsymbol{\lambda}$ and $\eta$ by the linear programming solver. These Lagrange multipliers correspond to the conversion of the three inequality constraints in \eqref{eq:multi-cuts-dw-master} into equality constraints and are collectively denoted as the vector $\bm{\pi}$.

Similarly, the programming \eqref{eq:single-cut} can be written as
\begin{equation}
    \begin{aligned}
        \min_{\boldsymbol{\rho}} \quad & \sum_{e=1}^{n_e} \sum_{m=1}^{n_M} \widetilde{w}_{e,m}^{k-1} \rho_{e, m} \\
        \text{s.t.} \quad & t^{k-1}(\bm{\rho}) \leq d^{k-1} \\
        & \sum_{e=1}^{n_e} \sum_{m=1}^{n_M} \frac{\bar{M}_m}{n_e} \rho_{e, m} \leq \bar{M}_{\max} \\
        & \sum_{m=1}^{n_M} \rho_{e, m} \leq 1, \quad e = 1, 2, \dots, n_e \\
        & \bm{\rho} \in \{0, 1\}^{n_e \times n_M} \; .
    \end{aligned}
    \label{eq:single-cut-dw-original}
\end{equation}
Applying the MDW decomposition to \eqref{eq:single-cut-dw-original}, the resultant $i$-th local sub-problem is given by: 
\begin{equation}
    \begin{aligned}
        \min_{\boldsymbol{\rho}_i} \quad & \sum_{e \in \mathcal{D}_i} \sum_{m=1}^{n_M} \widetilde{w}_{e,m}^{k-1} \rho_{e, m} + \pi^t t^{k-1}(\bm{\rho}_i) + \pi^M H_{M_0}(\bm{\rho}_i) \\
        \text{s.t.} \quad & \sum_{m=1}^{n_M} \rho_{e, m} \leq 1, \quad e \in \mathcal{D}_i \\
        & \bm{\rho}_i \in \{0, 1\}^{| \mathcal{D}_i | \times n_M}\; ,
    \end{aligned}
    \label{eq:single-cut-dw-sub}
\end{equation}
and the corresponding global sub-problem reads:
\begin{equation}
    \begin{aligned}
        \min_{\bm{\lambda}} \quad & \sum_{e=1}^{n_e} \left( \sum_{m=1}^{n_M} \left[ {\widetilde{w}_{e,m}^{k-1}} (\bar{\rho}_{e, m} - \rho^{k-1}_{e,m}) \right] \right) \\
        \text{s.t.} \quad & t^{k-1}(\bm{\bar{\rho}}) \leq d^{k-1} \\
        & \sum_{e=1}^{n_e} \sum_{m=1}^{n_M} \frac{\bar{M}_m}{n_e} \bar{\rho}_{e, m} \leq \bar{M}_{\max} \\
        & \sum_{q = 1}^{l} \lambda_{i}^q  = 1, \quad i = 1, 2, \dots, n_\mathcal{D} \\
        & \bm{\lambda} \in [0, 1]^{n_{\mathcal{D}} \times l} \;.
    \end{aligned}
    \label{eq:single-cut-dw-master}
\end{equation}
Like \eqref{eq:multi-cuts-dw-sub-reduced}, \eqref{eq:single-cut-dw-sub} is a binary programming problem, while \eqref{eq:single-cut-dw-master} is a linear programming problem, similar to  \eqref{eq:multi-cuts-dw-master}.

\subsubsection{Initialization}
To initiate the MDW decomposition, an initial guess of $\boldsymbol{\rho}$ to the programming \eqref{eq:branched_multi-cut} or \eqref{eq:single-cut} is necessary. Thus, we propose solving a reduced-size problem with a clustered design variable $\hat{\bm{\rho}} \in \{0, 1\}^{n_{\mathcal{D}} \times n_M}$ to obtain the initial guess. Specifically, we assign all design variables $\rho_{e, m}$ in the $i$-th block with a single value of $\hat{\rho}_{i, m}$, i.e., $\rho_{e, m} = \hat{\rho}_{i, m}, ~\forall e \in \mathcal{D}_i$. As such, the design variables $\rho_{e, m}$ belong to the $i$-th block are aggregated into one variable $\hat{\rho}_{i, m}$, and correspondingly the coefficients associated with the design variables in \eqref{eq:branched_multi-cut} or \eqref{eq:single-cut} are also aggregated. The resultant reduced-size multi-cut problem reads:
\begin{equation}
    \begin{aligned}
        \min_{\hat{\boldsymbol{\rho}}, \eta} \quad & \eta \\
        \text{s.t.} \quad & f(\bm{\rho}^j, \mathbf{u}^j) + \sum_{m=1}^{n_M} \sum_{i=1}^{n_{\mathcal{D}}} \left( \sum_{e \in \mathcal{D}_i} \widetilde{w}_{e,m}^j \right) \hat{\rho}_{i, m}  - \sum_{e=1}^{n_e} \sum_{m=1}^{n_M} \widetilde{w}_{e, m}^j \rho_{e, m}^j \leq \eta, \quad \forall j \in \mathcal{P}_s \\
        & \dfrac{1}{n_e} \sum_{m=1}^{n_M} \sum_{i=1}^{n_{\mathcal{D}}} \left[ \sum_{e \in \mathcal{D}_i} \left( 1 - 2 \sum_{m=1}^{n_M} \rho_{e, m}^j \right) \right] \hat{\rho}_{i, m} + \sum_{e=1}^{n_e} \left( \rho_{e, m}^j \right)^2 \leq d^j, \quad \forall j \in \mathcal{P}_s \\
        & \sum_{m=1}^{n_M} \sum_{i=1}^{n_{\mathcal{D}}} \frac{\bar{M}_m \cdot | \mathcal{D}_i |}{n_e} \hat{\rho}_{i, m} \leq \bar{M}_{\max} \\
        & \sum_{m=1}^{n_M} \hat{\rho}_{i, m} \leq 1, \quad i = 1, 2, \dots, n_{\mathcal{D}} \\
        & \hat{\bm{\rho}} \in \{0, 1\}^{n_{\mathcal{D}} \times n_M} \; .
    \end{aligned}
    \label{eq:multi-cuts-dw-reduced}
\end{equation}
And the resultant reduced-size single-cut problem is given by: 
\begin{equation}
    \begin{aligned}
        \min_{\hat{\boldsymbol{\rho}}} \quad & \sum_{m=1}^{n_M} \sum_{i=1}^{n_{\mathcal{D}}} \left( \sum_{e \in \mathcal{D}_i} \widetilde{w}_{e,m}^{k-1} \right) \hat{\rho}_{i, m} \\
        \text{s.t.} \quad & \dfrac{1}{n_e} \sum_{m=1}^{n_M} \sum_{i=1}^{n_{\mathcal{D}}} \left[ \sum_{e \in \mathcal{D}_i} \left( 1 - 2 \sum_{m=1}^{n_M} \rho_{e, m}^{k-1} \right) \right] \hat{\rho}_{i, m} + \sum_{e=1}^{n_e} \left( \rho_{e, m}^{k-1} \right)^2 \leq d^{k-1} \\
        & \sum_{m=1}^{n_M} \sum_{i=1}^{n_{\mathcal{D}}} \frac{\bar{M}_m \cdot | \mathcal{D}_i |}{n_e} \hat{\rho}_{i, m} \leq \bar{M}_{\max} \\
        & \sum_{m=1}^{n_M} \hat{\rho}_{i, m} \leq 1, \quad i = 1, 2, \dots, n_{\mathcal{D}} \\
        & \hat{\bm{\rho}} \in \{0, 1\}^{n_{\mathcal{D}} \times n_M} \; .
    \end{aligned}
    \label{eq:single-cut-dw-reduced}
\end{equation}
Both can be solved by a standard MILP solver. In case that the initialization problem \eqref{eq:multi-cuts-dw-reduced} or \eqref{eq:single-cut-dw-reduced} is infeasible, $n_{\mathcal{D}}$ will be doubled, \eqref{eq:multi-cuts-dw-reduced} or \eqref{eq:single-cut-dw-reduced} will then be reformulated with the new $n_{\mathcal{D}}$. This process can be repeated until a feasible solution to the initialization problem is found. \revthree{The rationale behind this process is that the reduced-sized problem imposes a restriction to the original problem as it constrains all design variables within a block to take the same value. Doubling the number of blocks, or equivalently halving the number of design variables within each block, thus relaxes this constraint and in turn increases the likelihood that the resulting problem is feasible.} Our numerical experiments indicate that infeasible solutions are uncommon during the initialization stage, provided $n_{\mathcal{D}}$ is not set too small. For all numerical experiments in this work, we set $n_{\mathcal{D}} = 100$. The solution to the initialization problem yields the initial Lagrange multiplier values, which are then used when solving the local sub-problems in the first iteration step.

Since the size of the clustered design variable $\hat{\bm{\rho}}$ depends only on the number of blocks ($n_{\mathcal{D}}$) and the number of materials ($n_M$), the initialization cost stays low, even when tackling large-scale TO problems.

\subsubsection{Iteration Process and Algorithm}
\label{subsubsec:initiation}
The iteration process of the MDW decomposition starts with the initialization step of solving the reduced-size problem \eqref{eq:multi-cuts-dw-reduced} or \eqref{eq:single-cut-dw-reduced}. Then at each iteration step $l$, the $n_D$ local sub-problems \eqref{eq:multi-cuts-dw-sub-reduced} or \eqref{eq:single-cut-dw-sub} are solved to yield $\rho^{k,l}_{e, m}$, with the Lagrangian multipliers obtained from solving the global sub-problem \eqref{eq:multi-cuts-dw-master} or \eqref{eq:single-cut-dw-master} in the last iteration. The design variables $\rho_{e, m}$ in the original programming \eqref{eq:branched_multi-cut} or \eqref{eq:single-cut} are finally updated by the design variables $\rho^{k,L}_{e, m}$ at the end of the iteration process of MDW.

The iteration process is terminated once the convergence criterion is satisfied, which is defined by the Lagrange multipliers $\bm{\pi}$ obtained from the global sub-problem, as:
\begin{equation}
    \dfrac{\left\| \bm{\pi}^l - \bm{\pi}^{l-1} \right\|_2}{\left\| \bm{\pi}^{l-1} \right\|_2} \leq 10^{-6} \;.    \label{eq:converg_criterion_Lagrange}
\end{equation}
Here, the vector $\bm{\pi}$ is defined as:  $\bm{\pi} =\{\pi^f, \pi^t, \pi^M\}$ if the original programming is multi-cut as in \eqref{eq:branched_multi-cut}, or $\bm{\pi} =\{\pi^t, \pi^M\}$ if the original programming is single-cut as in \eqref{eq:single-cut}. \revthree{The tolerance in Eq. \eqref{eq:converg_criterion_Lagrange} is consistently set to $10^{-6}$ throughout all numerical results presented in the main text. The robustness of the proposed method with respect to this parameter is discussed in \ref{sec:sens_tolerance}.} 

The entire process for the proposed MDW decomposition, including its initiation and iteration process, is summarized in Algorithm \ref{algorithm:dw_decomposition}.
\begin{algorithm}
    \caption{\textsc{MDW Decomposition}}\label{algorithm:dw_decomposition}
    \textbf{Input}: Programming \eqref{eq:branched_multi-cut} or \eqref{eq:single-cut} and the value of $n_{\mathcal{D}}$
    
    \textbf{Output}: The optimal solution $\bm{\rho}$ and the objective function value $\eta$ to the programming \eqref{eq:branched_multi-cut} or \eqref{eq:single-cut}
    
    \textbf{Utility}: Use MDW decomposition to solve the programming  \eqref{eq:branched_multi-cut} or \eqref{eq:single-cut}
    
    \begin{algorithmic}[1]
        \State Calculate $\widetilde{\bm{\rho}}$ according to \eqref{eq:rho_tilde} and sort $\widetilde{\bm{\rho}}$ in non-descending order
        \State Partition the design variables into $n_{\mathcal{D}}$ blocks according to \eqref{eq:dw-block}
        \State Solve the initialization problem \eqref{eq:multi-cuts-dw-reduced} or \eqref{eq:single-cut-dw-reduced}
        \If {\eqref{eq:multi-cuts-dw-reduced} or \eqref{eq:single-cut-dw-reduced}  is infeasible}
            \State $n_{\mathcal{D}} = 2 \cdot n_{\mathcal{D}}$
            \State Go to Step 2
        \Else
            \State Update the design variables $\rho_{e, m}^{k,0}$ with the solution of the initialization problem and update the corresponding Lagrange multipliers $\bm{\pi}^0$
        \EndIf
        \For {$l = 1, 2, \dots$}
            \State Solve all $n_D$ local sub-problems \eqref{eq:multi-cuts-dw-sub-reduced} or \eqref{eq:single-cut-dw-sub}
            \State Update the design variables $\rho_{e, m}^{k,l}$
            \State Solve the global sub-problem \eqref{eq:multi-cuts-dw-master} or \eqref{eq:single-cut-dw-master} and update the corresponding Lagrange multipliers $\bm{\pi}^l$
            \If {$\dfrac{\left\| \bm{\pi}^l - \bm{\pi}^{l-1} \right\|_2}{\left\| \bm{\pi}^{l-1} \right\|_2} \leq 10^{-6}$}
                \State \textbf{Break}
            \EndIf
        \EndFor
    \end{algorithmic}
    \hspace*{\algorithmicindent} \textbf{Return}: $\bm{\rho}^{k,l}$, $\eta$
\end{algorithm}

\subsection{Quantum Implementation}\label{subsec:QI}

A QUBO problem generally reads:
\begin{equation}
    \begin{aligned}        
    & \mathbf{\mathbb{Z}}^*=\min_{\mathbf{{\mathbb{Z}}}\in \{0, 1\}^n} \mathcal{J} \\ 
    & \mathcal{J}= \mathbf{{\mathbb{Z}}}^\intercal \mathbf{Q} \mathbf{{\mathbb{Z}}} = \sum_{i = 1}^n Q_{ii} \mathbb{Z}_i + \sum_{i = 1}^n \sum_{j = i + 1}^n Q_{ij} \mathbb{Z}_i \mathbb{Z}_j      
    \end{aligned} \;,
    \label{eq:QUBO_general}
\end{equation}
where $\mathbb{Z}$ denotes a binary variable, and $\mathbf{Q}$ represents the coefficient matrix. Directly converting the MILP problem \eqref{eq:branched_multi-cut} or \eqref{eq:single-cut} into an equivalent QUBO problem presents a significant hurdle. This is because when the global constraints that involve all design variables are incorporated into the objective function $\mathcal{J}$ (e.g., through a penalty method), they can lead to a dense coefficient matrix $\mathbf{Q}$. A dense $\mathbf{Q}$ matrix has two major implications. First, solving the resulting QUBO problem necessitates an all-to-all connection between qubits, which is a demanding requirement for current quantum hardware. Second, the computational cost of constructing the $\mathbf{Q}$ matrix itself scales quadratically with the number of design variables (i.e., $n_e \times n_M$). This substantial classical pre-computation effectively negates any potential quantum speedup, making the overall approach inefficient.

The proposed MDW decomposition introduced in \S\ref{sec:dw} provides a systematic way to decompose the MILP problems like \eqref{eq:branched_multi-cut} or \eqref{eq:single-cut} into a series of two types of sub-problems: Binary Programming (or Binary Integer Programming, BIP) \eqref{eq:multi-cuts-dw-sub-reduced} or \eqref{eq:single-cut-dw-sub}; and Linear Programming (LP) \eqref{eq:multi-cuts-dw-master} or \eqref{eq:single-cut-dw-master}. Thereby, we solve totally $n_\mathcal{D}$ BIP problems and one LP problem in each iteration step of the MDW decomposition. The LP problem can be easily solved by invoking an off-the-shelf classical LP solver, e.g., provided by Gurobi \cite{gurobi}. The overall computational cost is then dominated by solving $n_\mathcal{D}$ BIP problems. Compared to classical optimizers, quantum computing is known to be more efficient for solving BIP problems. For example, by leveraging the ``quantum tunneling'' effect, quantum annealing can pass through high energy peaks without requiring to follow the energy landscape, and hence effectively avoid falling into local minimum. As a result, it can find the global minimum (the lowest energy state) significantly faster. Therefore, this section focuses on the quantum implementation for solving the BIP problems \eqref{eq:multi-cuts-dw-sub-reduced} and \eqref{eq:single-cut-dw-sub}. 

To proceed, we formulate each problem into the form of QUBO. First, the binary constraints in both problems are the same and can be converted into the QUBO form as:
\begin{equation}
    \sum_{e \in \mathcal{D}_i} \left[ \sum_{m=1}^{n_M} \rho_{e, m} + \phi_e - 1 \right]^2 \;,
\end{equation}
where $\phi_e$ is a slackness variable introduced for restricting the material usage on each element $e$. 

Second, as the objective function is linear for both problems, the QUBO formed for each problem \eqref{eq:multi-cuts-dw-sub-reduced} or \eqref{eq:single-cut-dw-sub} requires only $(n_M + 1) \times |\mathcal{D}_i|$ logical qubits. While the QUBO for the multi-cut BIP problem \eqref{eq:multi-cuts-dw-sub-reduced} is given by:
\begin{equation}
    \begin{aligned}
        & \sum_{e \in \mathcal{D}_i} \sum_{j \in \mathcal{P}_s} \pi^f_j\widetilde{f}^j(\bm{\rho}_i) + \sum_{j \in \mathcal{P}_s} \pi_j^t t^j(\bm{\rho}_i) + \pi^M H_{M_0}(\bm{\rho}_i) 
        +  \Pi \sum_{e \in \mathcal{D}_i} \left[ \sum_{m=1}^{n_M} \rho_{e, m} + \phi_e - 1 \right]^2 \;,
    \end{aligned}
    \label{eq:multi-cuts-dw-qubo}
\end{equation}
the QUBO for the single-cut BIP problem \eqref{eq:single-cut-dw-sub} can be written as:
\begin{equation}
    \begin{aligned}
        & \sum_{e \in \mathcal{D}_i} \sum_{m=1}^{n_M} \widetilde{w}_{e,m}^{k-1} \rho_{e, m} + \pi^t t^{k-1}(\bm{\rho}_i) + \pi^M H_{M_0}(\bm{\rho}_i) 
        +  \Pi \sum_{e \in \mathcal{D}_i} \left[ \sum_{m=1}^{n_M} \rho_{e, m} + \phi_e - 1 \right]^2 \;.
    \end{aligned}
    \label{eq:single-cut-dw-qubo}
\end{equation}
Here, $\Pi$ in either of the above two equations is a large positive constant to penalize the binary constraint. As $\widetilde{f}^j$ in the cuts, $t^j$ for the trust regions, and the total mass $H_{M_0}$ are all bounded with respect to the design variables $\bm{\rho}$, $\Pi$ can be straightforwardly set to their upper bounds as:
\begin{equation}
    \Pi = \max \left( \pi^f_j \sum_{e \in \mathcal{D}_i} \sum_{m=1}^{n_M} |\widetilde{w}_{e,m}^j|, \ \pi_j^t d^j, \ \pi^M \bar{M}_{\max} \right) 
\end{equation}
for \eqref{eq:multi-cuts-dw-sub-reduced} and
\begin{equation}
    \Pi = \max \left( \sum_{e \in \mathcal{D}_i} \sum_{m=1}^{n_M} |\widetilde{w}_{e,m}^{k-1}|, \ \pi^t d^{k-1}, \ \pi^M \bar{M}_{\max} \right) 
\end{equation}
for \eqref{eq:single-cut-dw-sub}. Here, $\sum_{e \in \mathcal{D}_i}\sum_{m=1}^{n_M} |\widetilde{w}_{e,m}^j|$ is used as the upper bound of $\widetilde{f}^j(\bm{\rho}_i)$.

\section{Results}\label{sec:results}
The proposed method is evaluated for its solution quality and computational efficiency through large-scale 3D bridge design problems, including both single and multi-material cases. Robustness and scalability are assessed by varying the discretization resolution, with the largest design comprising over 10 million discrete elements and more than 50 million design variables. Comparative analyses against representative classical methods, including SIMP~\cite{jia2024fenitop} and DVTO-MT \cite{ye2025discrete}, demonstrate the superior performance of our new method.

The FEM solver used in this work is \texttt{FEniCSx}~\cite{BarattaEtal2023}, which is based on \texttt{FEniTop}~\cite{jia2024fenitop}, a parallel scalable implementation that employs \texttt{FEniCSx} for solving TO problems.

\subsection{Problem Setup}
\begin{figure}[htp!]
    \centering
    \includegraphics[width=0.48\textwidth]{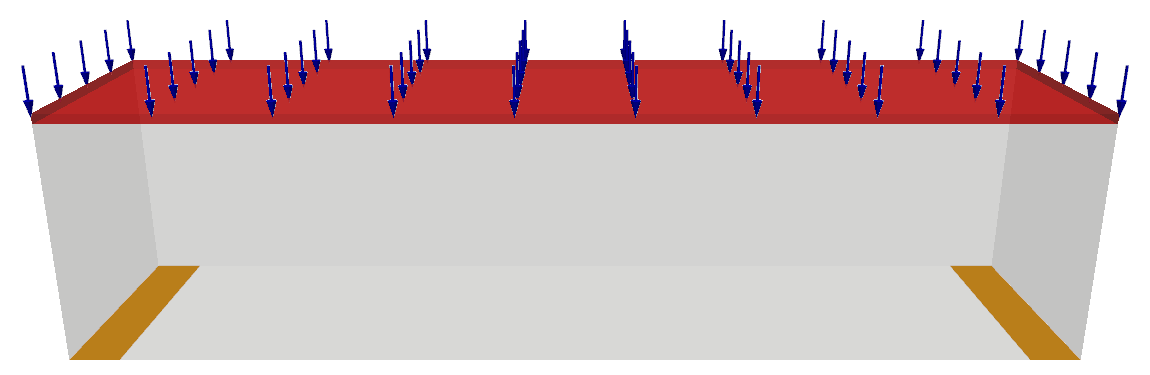}
    \caption{Domain setup and boundary conditions for 3D bridge design.}\label{fig:bridge}
\end{figure}

The design domain is a 3D box, with the length of $L = 40$ m, the width of $W = 10$ m, and the height of $H = 10$ m, as depicted in \Cref{fig:bridge}. A solid plate with the thickness of 0.4 m is placed on top of the box, designated by the red region in \Cref{fig:bridge}. This top plate region is excluded from the TO design, and hence is always assigned solid elements of the strongest material. The design domain is fixed in two striped regions at either end of the bottom boundary, each with a width of 2 m, as designated by the orange regions in \Cref{fig:bridge}. The gray region represents the actual design domain, allowing for free-form topologies and the use of any candidate materials, as determined by the optimization solution. A uniformly distributed loading force with the magnitude of $-10^3 N/m^2$ is exerted on the top boundary of the design domain, as indicated by blue arrows in \Cref{fig:bridge}. Three different discretization resolutions are examined, including $50 \times 200 \times 50$, $100 \times 400 \times 100$, and $150 \times 600 \times 150$. 

For the single-material design scenario, the material considered is stainless steel. Whereas, for multi-material TO designs, we include four representative metallic materials, whose mechanical properties are detailed in \Cref{tab:four-metallic-material-properties}. 
These materials, magnesium, aluminum, titanium, and stainless steel, represent a selection of the most widely utilized metallic materials in structural engineering applications. They also span a range of properties, including Young's modulus, Poisson's ratio, and density. The inclusion of diverse materials allows us to demonstrate the proposed method's capability to autonomously select optimal materials during optimization. For the multi-material design, we set the maximum total mass to $3.744\times 10^6$ kg. This value is equivalent to a 12\% volume fraction if only steel were used in the single-material design. 
Given that the top plate of the design domain consistently uses steel across all scenarios,
the mass of material actually subject to optimization is reduced to $2.496\times 10^6$ kg for the multi-material design, which corresponds to an 8\% volume fraction when only using steel for the single-material design. The design objective is to minimize the compliance of the bridge, i.e., in Eq. \eqref{eq:non-convex-to}, $f(\mathbf{u}, \bm{\rho}) = \mathbf{f}^\intercal \mathbf{u}$. With this design, the bridge's topology and usage of candidate material(s) are optimized to withstand the applied external load. 
\begin{table*}[htbp]
    \centering
    \caption{Properties of the four candidate materials considered in multi-material TO design.}\label{tab:four-metallic-material-properties}
    \begin{tabular}{ccccc}
        \toprule
        Materials & \textbf{Magnesium} & \textbf{Aluminum} & \textbf{Titanium} & \textbf{Stainless steel} \\
        \midrule
        Young's modulus $E_m \ (\text{GPa})$ & 44 & 73 & 100 & 210 \\
        Poisson's ratio $\nu_m$ & 0.28 & 0.33 & 0.36 & 0.29 \\
        Mass density $\bar{M}_m {~(10^3 \times \text{kg}/\text{m}^3)}$ & 1.74 & 2.70 & 4.50 & 7.80 \\
        Specific strength $\frac{E_m}{\bar{M}_m} {~(10^6 \times \frac{N\cdot\text{m}}{\text{kg}})}$ & 25.3 & 27.0  & 22.2 & 26.9\\
        \bottomrule
    \end{tabular}
\end{table*}

\subsection{Single-Material Design}\label{subsec:results-single}
For the single-material design, $n_M = 1$, and as mentioned above, stainless steel is the material used for the design. The results obtained from three different methods are compared, including the proposed MDW decomposition method in this work, the SIMP method \cite{jia2024fenitop}, and the original DVTO-MT method \cite{ye2025discrete}. In light of current quantum hardware access limitations, we validate our proposed MDW decomposition method through a two-tiered approach. The first tier involves validating its classical implementation by solving local sub-problems \eqref{eq:multi-cuts-dw-sub-reduced} or \eqref{eq:single-cut-dw-sub} with a classical optimizer provided by Gurobi \cite{gurobi}. The corresponding results are referred to as ``\textbf{MDW-CC}". In the second tier, we validate the QUBO formulation for the multi-cut BIP problem \eqref{eq:multi-cuts-dw-sub-reduced} by solving it using the QAOA algorithm.

For the first-tier validation, we begin with comparing the results with those of SIMP. \Cref{tab:single-mat-dw-simp-obj} summaries the results for the optimal objective function value ($f$), 
the total iteration count ($N_{\text{Iter}}$), and the total computing time ($T_{\text{tot}}$, at different discretization resolutions. The total iteration count for solving the TO problem \eqref{eq:non-convex-to} dictates the number of FEM solver calls, making it a key indicator of methodological efficiency, alongside computing time. For the SIMP results, we adopted a recent scalable implementation, \textit{FEniTop}~\cite{jia2024fenitop}, used the optimality criterion (OC) for its optimal performance, and set the maximum number of iterations to 400 and the Helmholtz filter radius to 0.6. By comparing the objective function values, we find that the SIMP method requires at least 400 iterations to achieve comparably low values; terminating earlier, such as at 200 iteration steps, does not yield sufficiently low objective function values. This significantly higher iteration count results in far more frequent FEM solution invocations and a greatly increased number of optimization problem solutions. This, in turn, leads to much longer total computing times for SIMP, almost an order of magnitude greater than our method's classical implementation (MDW-CC). 
Moreover, since the $n_D$ local sub-problems can be solved independently, our implementation solves them in parallel during each MDW iteration. \revtwo{All computing times reported herein are wall times measured on a single node equipped with two AMD EPYC 7763 64-Core Processors and 512 GB of memory. Each test was launched via the Slurm system \cite{yoo2003slurm}, utilizing all 128 physical cores of the node.}
\begin{table*}[hbt!]
    \caption{Single-material design: The resultant objective function value ($f$), total iteration count ($N_{\text{Iter}}$), and total computing time ($T_\text{tot}$), compared with the SIMP method \cite{jia2024fenitop}.}\label{tab:single-mat-dw-simp-obj}
    \centering
    \begin{tabular}{c|ccc|ccc|ccc}
        \toprule
        \multirow{2}{*}{\textbf{Resolution}} & \multicolumn{3}{c|}{\textbf{MDW-CC}} & \multicolumn{3}{c|}{\textbf{SIMP ($N_\text{Iter} = 200$)}} & \multicolumn{3}{c}{\textbf{SIMP ($N_{\text{Iter}} = 400$)}} \\
        \cmidrule{2-4} \cmidrule{5-7} \cmidrule{8-10}
        & $f$ & $N_{\text{Iter}}$ & $T_\text{tot}$ (s) & $f$ & $\Delta f$ (\%) & $T_\text{tot}$ (s) & $f$ & $\Delta f$ (\%) & $T_\text{tot}$ (s) \\
        \midrule
        $50 \times 200 \times 50$ & $15.229$ & $77$ & $370.9$ & $19.172$ & $25.9$ & $1015.4$ & $16.148$ & $6.0$ & $2821.8$\\
        $100 \times 400 \times 100$ & $14.115$ & $67$ & $2445.7$ & $18.418$ & $30.5$ & $9235.7$& $14.499$ & $2.7$ & $20261.4$\\
        $150 \times 600 \times 150$ & $13.205$ & $47$ & $6434.2$ & $18.231$ & $38.1$ & $37004.4$ & $14.048$ & $6.4$ & $84163.0$\\
        \bottomrule
    \end{tabular}
\end{table*}

\begin{figure}[htp]
    \centering
    \begin{subfigure}{.6\textwidth}
        \centering
        \includegraphics[width=\textwidth]{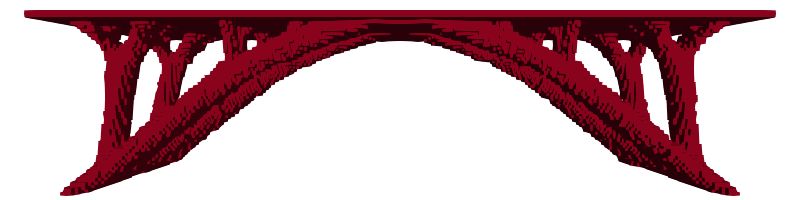}
        \caption{Discretization resolution of $50 \times 200 \times 50$.}
    \end{subfigure}
    \begin{subfigure}{.6\textwidth}
        \centering
        \includegraphics[width=\textwidth]{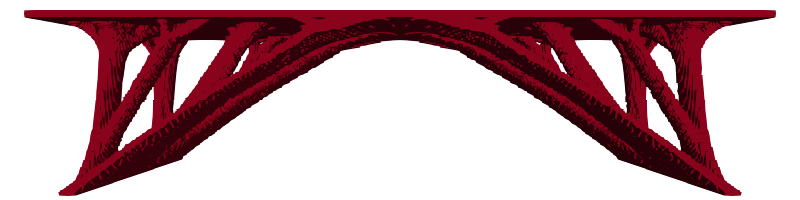}
        \caption{Discretization resolution of $100 \times 400 \times 100$.}
    \end{subfigure}
    \begin{subfigure}{.6\textwidth}
        \centering
        \includegraphics[width=\textwidth]{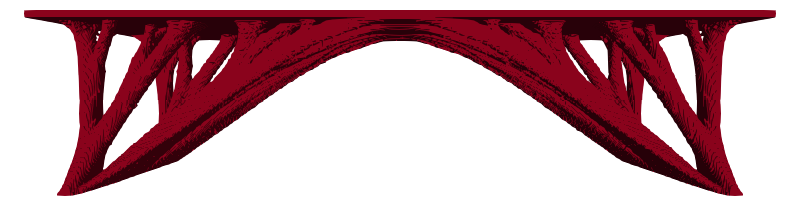}
        \caption{Discretization resolution of $150 \times 600 \times 150$.}
    \end{subfigure}
    \caption{Single-material design: The optimal topologies yield by our method with three discretization resolutions.}\label{fig:dw-single-topology}
\end{figure}

\begin{figure}[htp]
    \centering
    \begin{subfigure}{.6\textwidth}
        \centering
        \includegraphics[width=\textwidth]{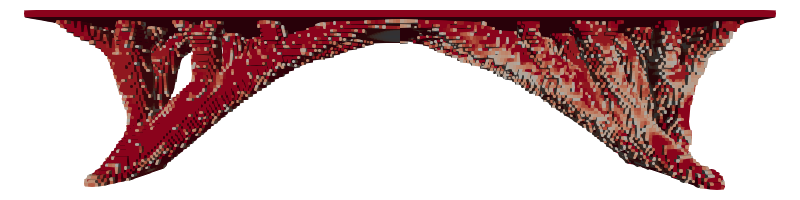}
        \caption{Discretization resolution of $50 \times 200 \times 50$.}
    \end{subfigure}
    \begin{subfigure}{.6\textwidth}
        \centering
        \includegraphics[width=\textwidth]{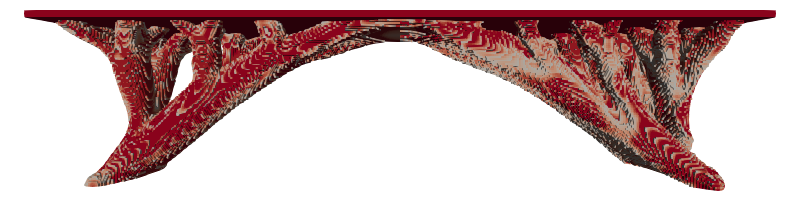}
        \caption{Discretization resolution of $100 \times 400 \times 100$.}
    \end{subfigure}
    \begin{subfigure}{.6\textwidth}
        \centering
        \includegraphics[width=\textwidth]{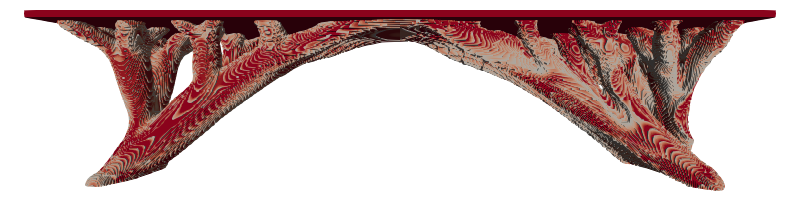}
        \caption{Discretization resolution of $150 \times 600 \times 150$.}
    \end{subfigure}
    \caption{Single-material design: The optimal topologies produced by the SIMP method with three discretization resolutions. For each topology, the left half presents the outcome after 200 iteration steps, and the right half after 400 iteration steps. The color intensity reflects the values of the design variable $\rho$ following a post-processing filter with the threshold of 0.5, where darker shades indicate values closer to 1 and lighter shades represent values closer to 0.5.}\label{fig:simp-topology}
\end{figure}

Next, we compare the resultant optimal topologies. The ones obtained by MDW-CC are shown in \Cref{fig:dw-single-topology} at varying discretization resolutions. Our method can generate clear-cut 0/1 topologies, because it directly uses binary design variables. As the discretization is refined, the optimal topology reveals finer structures and develops more branches under the top plate. 
The optimal topologies obtained by the SIMP method are reported in \Cref{fig:simp-topology}. Since SIMP does not produce clear-cut topologies, the results in \Cref{fig:simp-topology} are shown after a post-processing filter with the threshold of 0.5, by which the design variable $\rho$ ranges continuously from 0.5 to 1, as depicted by varied colors in \Cref{fig:simp-topology}. Comparing the left half (presenting the outcome after 200 iteration steps) and the right half (after 400 iteration steps), it is evident that SIMP needs more iterations to reach not only lower objective function values (as indicated in \Cref{tab:single-mat-dw-simp-obj}) but also further evolving topologies.

Further comparison with the original DVTO-MT method \cite{ye2025discrete} focuses solely on the total time dedicated to solving master problems, denoted as $T_\text{OPT}$, because MDW also adheres to the DVTO-MT framework, solving the same primal problem and employing the same parameter relaxation scheme, thus requiring the same total iteration counts ($N_{\text{Iter}}$) and FEM solution time ($T_{\text{FEM}}$) as the original DVTO-MT method \cite{ye2025discrete}. This comparison, detailed in \Cref{tab:single-mat-dw-milp-opt}, elucidates the superior efficiency of MDW compared to employing a MILP solver \cite{gurobi} for master problems \eqref{eq:branched_multi-cut} or \eqref{eq:single-cut}. The optimization time incurred by the MILP solver is comparable to FEM solution time, making the solution of large-scale TO problems very expensive. In contrast, our MDW method achieves substantial speedup, reaching more than an order of magnitude. This speedup becomes more pronounced with increasing problem size, establishing MDW as a practically efficient approach for 3D large-scale TO problems. For multi-material designs, the speedup proves even more significant, a point elaborated later in \S\ref{subsec:results-multi}.
\begin{table*}[hbt!]
    \caption{Single-material design: The time spent for optimization ($T_\text{OPT}$), compared with the original DVTO-MT method \cite{ye2025discrete}. Here, the FEM solution time ($T_{\text{FEM}}$) is provided for reference.}\label{tab:single-mat-dw-milp-opt}
    \centering
    \begin{tabular}{c|cccc}
        \toprule
        \textbf{Resolution} & $T_{\text{FEM}}$ (s) & $T_\text{OPT}^\text{MDW-CC}$ (s) & $T_\text{OPT}^\text{DVTO-MT}$ (s) & $T_\text{OPT}^\text{DVTO-MT}/T_\text{OPT}^\text{MDW-CC}$ \\
        \midrule
        $50 \times 200 \times 50$ & $303.3$ & $67.6$ & $268.1$ & $4.0$ \\
        $100 \times 400 \times 100$ & $2282.2$ & $163.5$ & $2824.8$ & $17.3$ \\
        $150 \times 600 \times 150$ & $6234.0$ & $200.2$ & $5254.6$ & $26.2$ \\
        \bottomrule
    \end{tabular}
\end{table*}

As discussed in \S\ref{subsec:QI}, the local sub-problems \eqref{eq:multi-cuts-dw-sub} or \eqref{eq:single-cut-dw-sub} can be solved using quantum computing to leverage its advantages in solving BIP problems. Thus, in the second-tier validation, we verify the QUBO formulated for the local sub-problems. Due to the practical limitations of accessing real quantum hardware, we opted to solve the QUBO in \eqref{eq:multi-cuts-dw-qubo} with the QAOA algorithm, leveraging the simulator provided by \texttt{CUDA Quantum} \cite{CUDA-Q}. \revtwo{The setup of QAOA requires a parameterized quantum circuit (ansatz) with a prescribed depth $\mathcal{L}$ and a Hamiltonian $\boldsymbol{\mathcal{H}}$ for defining the objective function. More specifically, following the procedure described in \cite{MaxCut-QAOA}, a quantum state $\ket{\mathbf{\Xi}}$ was prepared by a parameterized quantum circuit with the prescribed depth $\mathcal{L}$. To construct the Hamiltonian $\boldsymbol{\mathcal{H}}$, we converted the QUBO to the Ising model through the following steps. Considering the QUBO in Eq. \eqref{eq:QUBO_general}, $\mathbb{Z}_i$ can be mapped into $\zeta_i$ by: 
\begin{equation}
    \mathbb{Z}_i = \dfrac{\zeta_i + 1}2, \quad \mathbb{Z}_i \in \{0, 1\}, \quad \zeta_i \in \{-1, 1\} \;.
\end{equation}
The objective function $\mathcal{J}$ can then be rewritten as:
\begin{equation}
    \mathcal{J} = \sum_{i = 1}^{n} Q_{ii} \left(\dfrac{\zeta_i + 1}2\right) + \sum_{i=1}^n\sum_{j=i+1}^n Q_{ij} \left(\dfrac{\zeta_i + 1}2\right) \left(\dfrac{\zeta_j + 1}2\right) ~~ \to ~~ \mathcal{J}^\prime = \sum_{i=1}^n \dfrac12\left({Q_{ii} + \sum_{j=i+1}^n Q_{ij}}\right) \zeta_i + \sum_{i=1}^n \sum_{j=i+1}^n \frac14 Q_{ij} \zeta_i \zeta_j
\end{equation}
where $\mathcal{J}^\prime$ omits the constants in $\mathcal{J}$. Thus, we obtained the Hamiltonian as:
\begin{equation}
    \boldsymbol{\mathcal{H}} = \sum_{i=1}^n \dfrac12\left({Q_{ii} + \sum_{j=i+1}^n Q_{ij}}\right) \mathbf{Z}_i + \sum_{i=1}^n \sum_{j=i+1}^n \frac14 Q_{ij} \mathbf{Z}_i \mathbf{Z}_j \;,
\end{equation}
where $\mathbf{Z}_i$ denotes the the Pauli-Z operator acting on the $i$-th qubit. The objective function for QAOA is finally defined as $\bra{\bm{\Xi}} \boldsymbol{\mathcal{H}} \ket{\bm{\Xi}}$.}

Given the limited number of qubits that can be simulated on a classical computer, we constrained the size of $|\mathcal{D}_i|$ up to 8 for validation purposes. 
\revtwo{Thirty local sub-problems were randomly selected from the case with the discretization resolution of $50\times 200 \times 50$. The number of layers in the ansatz was set to $\mathcal{L}= 4$ or $\mathcal{L}= 5$.} To evaluate accuracy, we compared the QUBO's solution generated by QAOA against that yield by calling the classical optimizer \cite{gurobi} for solving the original local sub-problem. The accuracy is evaluated by the percentage of agreement between the two solutions across all 30 problems solved, as presented in \Cref{Tab:QAOA}. The results show that with a sufficient number of layers in the ansatz, specifically $\mathcal{L}= 5$, our QUBO formulation successfully guides the optimization to the desired solution, thus validating its use for updating design variables in the TO process.
\begin{table*}[hbt!]
    \caption{Validation of QUBO: \eqref{eq:multi-cuts-dw-qubo} solved by QAOA for 30 randomly selected local sub-problems. Here, the accuracy is evaluated by the percentage of agreement with the solutions yield by calling the classical optimizer \cite{gurobi} for solving \eqref{eq:multi-cuts-dw-sub-reduced};  $\mathcal{L}$ denotes the number of layers in the ansatz; and $|\mathcal{D}_i|$ is the sub-problem's size.}
    \centering
    \begin{tabular}{c|cccccc}
        \toprule
        \multicolumn{7}{c}{$\mathcal{L} = 4$} \\
        \midrule
        $| \mathcal{D}_i |$ & 3 & 4 & 5 & 6 & 7 & 8 \\
        Accuracy & $100\%$ & $100\%$ & $100\%$ & $98.3\%$ & $94.3\%$ & $64.2\%$ \\
        \midrule
        \multicolumn{7}{c}{$\mathcal{L} = 5$} \\
        \midrule
        $| \mathcal{D}_i |$ & 3 & 4 & 5 & 6 & 7 & 8 \\
        Accuracy & $100\%$ & $100\%$ & $100\%$ & $100\%$ & $100\%$ & $95.4\%$ \\
        \bottomrule
    \end{tabular}
    \label{Tab:QAOA}
\end{table*}

Finally, we assess the potential quantum speedup. As previously mentioned, quantum computing can accelerate the solution of local sub-problems, which is the dominant cost in solving the master problem \eqref{eq:branched_multi-cut} or \eqref{eq:single-cut} via MDW. Although we cannot access actual quantum hardware to solve them, we can estimate the computing time and potential quantum speedup. The expected computing time on a QPU, when quantum annealers \cite{dwave-documentation} are used to solve the QUBO in \eqref{eq:multi-cuts-dw-qubo}, can be estimated as: 
\begin{equation}
    T_\text{QPU} = 20\mu s \times 1000 \times N_\text{Local} \;,
\end{equation}
where $T_\text{QPU}$ denotes the expected total QPU time for 1000 repetitions of annealing, with each annealing process taking 20$\mu s$ \cite{dwave-documentation}; and $N_\text{Local}=N_\text{Iter}\times L$ is the total number of local sub-problems involved in the entire solution process, which can be evaluated from our classical implementation of MDW. \revone{The embedding overhead on the QPU is excluded from this estimation. This is because embedding time is difficult to quantify in the absence of access to actual quantum hardware. However, it is worth emphasizing that a key advantage of the proposed MDW decomposition is that the resulting QUBOs \eqref{eq:multi-cuts-dw-qubo} or \eqref{eq:single-cut-dw-qubo} yield \textit{sparse} Hamiltonians that do not require all-to-all qubit connectivity, thereby avoiding the long embedding times typically associated with dense Hamiltonians, a critical consideration for near-term quantum hardware.} It should also be noted that, since the $n_D$ local sub-problems can be solved independently in each MDW iteration and were solved in parallel in our classical implementation, for a fair comparison, we assume that the quantum hardware accessed can simultaneously solve the corresponding $n_D$ QUBO problems. Such estimated time is presented in \Cref{tab:single-mat-qc-speedup}. \revone{In addition to the QPU time, the actual cost of solving the QUBO must also account for the time spent on forming the QUBO \eqref{eq:multi-cuts-dw-qubo} or \eqref{eq:single-cut-dw-qubo}, denoted as $T_\text{QUBO}$. This step, performed on a classical computer, was implemented with its wall time measured and reported in \Cref{tab:single-mat-qc-speedup}.} Thus, the potential quantum speedup factor is estimated from: 
\begin{equation}
    \text{Speedup} = \frac{T_\text{Local}^\text{MDW-CC}}{T_\text{QPU}+T_\text{QUBO}} \;,
\end{equation}
where $T_\text{Local}^\text{MDW-CC}$ denotes the total wall time incurred by using the classical optimizer to solve all local sub-problems. The speedup, as shown in \Cref{tab:single-mat-qc-speedup}, becomes more pronounced with larger problem sizes. This speedup can be more significant for multi-material designs, a point discussed further in \S\ref{subsec:results-multi}.
\begin{table*}[hbt!]
    \caption{Single-material design: Quantum computing time estimated and potential speedup.}\label{tab:single-mat-qc-speedup}
    \centering
    \begin{tabular}{c|ccccc}
        \toprule
        \textbf{Resolution} & $N_\text{Local}$ & $T_\text{QPU}$ (s) & $T_\text{QUBO}$ (s) & $T_\text{Local}^\text{MDW-CC}$ (s) & Speedup \\
        \midrule
        $50 \times 200 \times 50$ & $1039$ & $20.8$  & $0.1$  & $29.0$ & $1.4$ \\
        $100 \times 400 \times 100$ & $1444$ & $28.9$ & $1.5$ & $90.3$ & $3.0$ \\
        $150 \times 600 \times 150$ & $699$ & $14.0$ & $1.7$ & $137.8$ & $8.8$ \\
        \bottomrule
    \end{tabular}
\end{table*}

\subsection{Multi-Material Design}\label{subsec:results-multi}
\begin{figure}[htp]
    \centering
    \begin{subfigure}{.6\textwidth}
        \centering
        \includegraphics[width=\textwidth]{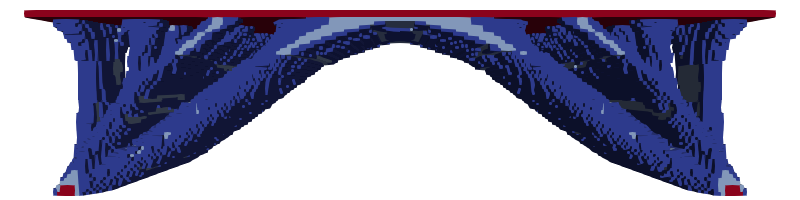}
        \caption{Discretization resolution of $50 \times 200 \times 50$}
    \end{subfigure}
    \begin{subfigure}{.6\textwidth}
        \centering
        \includegraphics[width=\textwidth]{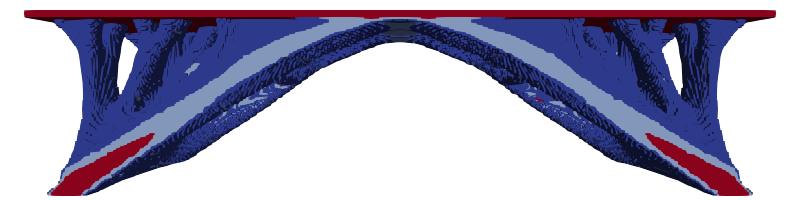}
        \caption{Discretization resolution of $100 \times 400 \times 100$}
    \end{subfigure}
    \begin{subfigure}{.6\textwidth}
        \centering
        \includegraphics[width=\textwidth]{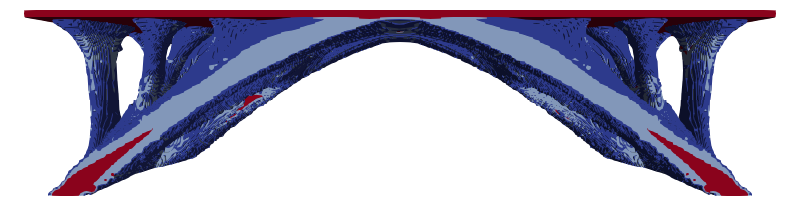}
        \caption{Discretization resolution of $150 \times 600 \times 150$}
    \end{subfigure}
    \caption{Multi-material design: The optimal topologies and material selections obtained from our method at three discretization resolutions. Materials are rendered as follows: Magnesium (dark blue), Aluminum (light blue), Titanium (orange), and Stainless Steel (red).}\label{fig:dw-multi-topology}
\end{figure}

For the multi-material design, all four candidate materials in \Cref{tab:four-metallic-material-properties} are involved. The design goal is twofold: optimizing the bridge's structural topology and selecting the optimal material combination, as shown in \Cref{fig:dw-multi-topology}.
Among the four candidate materials, only three were selected in the final design: titanium was excluded. This is because titanium's significantly lower specific strength ($E_m / \bar{M}_m$) makes it less efficient for minimizing compliance while maintaining a lightweight structure. The other three materials have comparable specific strengths, but stainless steel, despite its high Young's modulus, has the greatest mass density. As a result, it forms the smallest fraction of the final material selection, primarily used for parts near the center beneath the plate and close to the two bottom supports where maximum stiffness is needed. Both magnesium and aluminum are lightweight and thus dominate the final material selection. Magnesium, being the softest, appears mostly as the outer coating, while aluminum serves as an intermediate, transitional material between magnesium and stainless steel. This alignment with intuitive material selection principles supports our method's ability to correctly identify optimal materials.

Since the SIMP method \cite{jia2024fenitop} cannot account for multiple materials, we only compare our method with the original DVTO-MT method \cite{ye2025discrete} herein. Similar to the single-material design, our comparison with the original DVTO-MT method \cite{ye2025discrete} focuses solely on $T_\text{OPT}$, the total time dedicated to solving master problems \eqref{eq:branched_multi-cut} or \eqref{eq:single-cut}, as detailed in \Cref{tab:multi-mat-dw-milp-opt}. While both methods maintain the low iteration counts and comparable FEM solution time as observed in the single-material case, the multi-material design involves four times the number of design variables. This increase of design variables escalates the time spent solving the master problems compared to the single-material scenario. Notably, MDW's cost increases only slightly, whereas the MILP solver's cost rises exponentially, causing its optimization time to even exceed the FEM solution time by orders of magnitude. For the case of $100\times 400 \times 100$, our method achieves a remarkable speedup by three orders of magnitude. Furthermore, for the largest problem ($150\times 600 \times 150$, resulting in 54 million design variables), the MILP solver fails to provide a solution, yet MDW's time remains less than the FEM solution time, further substantiating its superior efficiency and practical utility for tackling large-scale 3D TO problems.
\begin{table*}[hbt!]
    \caption{Multi-material design: The resultant objective function value ($f$), total iteration count ($N_{\text{Iter}}$), and the FEM solution time ($T_{\text{FEM}}$). The time spent for optimization ($T_\text{OPT}$) is compared with the original DVTO-MT method \cite{ye2025discrete}.}\label{tab:multi-mat-dw-milp-opt}
    \centering
    \begin{tabular}{c|cccccc}
        \toprule
        \textbf{Resolution} & $f$ & $N_{\text{Iter}}$ & $T_{\text{FEM}}$ (s) & $T_\text{OPT}^\text{MDW-CC}$ (s)  &$T_\text{OPT}^\text{DVTO-MT}$ (s) & $T_\text{OPT}^\text{DVTO-MT}/T_\text{OPT}^\text{MDW-CC}$\\
        \midrule
        $50 \times 200 \times 50$ & $12.957$ & $39$ & $123.6$ & $61.1$ & $6118.6$ & $100$ \\
        $100 \times 400 \times 100$ & $13.052$ & $44$ & $1746.6$ & $285.2$ & $349929.0$ & $1227$ \\
        $150 \times 600 \times 150$ & $12.983$ & $51$ &  $9667.1$ & $2089.3$ & $-$ & $-$ \\
        \bottomrule
    \end{tabular}
\end{table*}

The assessment of potential quantum speedup for multi-material designs follows the same approach as in the single-material cases. \Cref{tab:multi-mat-qc-speedup} summarizes the estimated speedup. We observe that the speedup becomes increasingly pronounced as the problem size grows. For the largest test case, with a discretization resolution of $150 \times 600 \times 150$, the observed rise in $T_{\text{QUBO}}$, i.e., the time required to construct the QUBO, primarily results from an inefficient Python implementation relying on nested loops. This bottleneck can be substantially mitigated through more efficient coding practices, such as vectorizing the Python implementation or transitioning to a compiled language like C++. Such improvements are expected to substantially enhance the scalability of $T_{\text{QUBO}}$ and greatly reduce it for larger problems. This will, in turn, make the quantum speedup even more pronounced with increasing problem sizes. It is also worth noting that quantum advantage is significantly greater for multi-material designs compared to single-material designs. Thus, we anticipate that quantum computing will become increasingly impactful as the complexity and scale of TO designs continue to grow in real-world applications.
\begin{table*}[hbt!]
    \caption{Multi-material design: Quantum computing time estimated and potential speedup.}\label{tab:multi-mat-qc-speedup}
    \centering
    \begin{tabular}{c|ccccc}
        \toprule
        \textbf{Resolution} & $N_\text{Local}$ & $T_\text{QPU}$ (s) & $T_\text{QUBO}$ (s) & $T_\text{Local}^\text{MDW-CC}$ (s) & Speedup \\
        \midrule
        $50 \times 200 \times 50$ & 659 & 13.2  & 0.3  & 30.3 & 2.3 \\
        $100 \times 400 \times 100$ & 756 & 15.1 & 2.2 & 224.7 & 13.0 \\
        $150 \times 600 \times 150$ & 1400 & 28.0 & 29.7 & 1626.1 & 28.2 \\
        \bottomrule
    \end{tabular}
\end{table*}

\section{Conclusion}\label{sec:conclusion}
Two key factors largely determine the total computational time in TO: (1) the number of iterations required to achieve convergence, which dictates the number of PDE solver (e.g., FEM) calls, and (2) the time needed to solve the optimization problem at each iteration. The DVTO-MT framework developed in our prior work \cite{ye2025discrete} significantly reduces the number of iterations, and consequently, the time spent solving the governing PDE via FEM. Building upon this framework, the MDW decomposition proposed in the present work further substantially reduces the time needed to solve the optimization problem at each iteration. Accordingly, a major contribution of this work is the development of an efficient TO methodology that markedly enhances computational efficiency by reducing both the iteration count and per-iteration optimization time. This work has targeted on addressing large-scale, multi-material TO challenges for 3D continuum structures.

Another key contribution lies in enabling the integration of quantum computing to leverage potential quantum advantages. The MDW decomposition partitions each MILP problem into $n_D$ local sub-problems (BIPs) and one global sub-problem (LP). The BIP sub-problems, being the most computationally intensive, can be accelerated via quantum computing by solving their equivalent QUBO formulations. The formulated QUBOs are compatible with both quantum annealing and gate-based quantum computing. This represents a significant advancement over previous efforts that applied quantum computing to TO \revtwo{\cite{wang2024mapping,honda2024development,TO_DA_IEEE2022,sato2023quantum,kim2023topology,xiao2025efficient,ye2023quantum,sukulthanasorn2025novel,wang2025quantum,kim2025variational}, which primarily focused on discrete truss or small-scale continuum problems and were limited to single-material designs}.

We have evaluated the computational efficiency and solution quality of our method by designing 3D bridges using both single and multiple materials. As it directly handles binary design variables, our method naturally produces clear-cut 0/1 topologies, unlike SIMP \cite{jia2024fenitop}, the predominant TO approach, which relaxes binary variables into continuous ones (varying between 0 and 1) and requires post-processing, such as Heaviside projection \cite{kawamoto2011heaviside}, to revert to binary representations. In addition, our method achieves comparable objective function values while reducing overall computation time by at least one order of magnitude compared to SIMP \cite{jia2024fenitop}. When benchmarked against the original DVTO-MT \cite{ye2025discrete} on the time required for solving the master problems, the present method with the MDW decomposition can achieve speedups from one to three orders of magnitude. And it maintains low computational time even for the largest problem tested ($150\times 600 \times 150$, yielding 54 million design variables), for which the classical MILP solver fails to return a solution. The proposed quantum implementation has the potential to further accelerate solving the resultant BIP sub-problems by about another order of magnitude. Due to current constraints in accessing real quantum hardware, we have verified the validity of the formulated QUBO using QAOA \cite{CUDA-Q}. All the speedups noted above consistently become more pronounced with increasing problem size and complexity (moving from single-material to multi-material designs). This suggests that the new method and quantum computing will play an increasingly valuable role in addressing the scale and complexity of real-world TO applications.

Another important advantage of the proposed MDW decomposition is that each resulting BIP (local sub-problem) involves only local constraints. As a result, the corresponding QUBO formulation features a sparse Hamiltonian, eliminating the need for all-to-all qubit connectivity, a key consideration for near-term quantum hardware. In this regard, the proposed approach surpasses previous studies, including our earlier work \cite{ye2023quantum} and others \cite{sukulthanasorn2025novel}, which produced QUBOs with dense Hamiltonians requiring all-to-all connectivity and incurring quadratically scaled construction costs with respect to the number of design variables. Moreover, since all local sub-problems are independent within each MDW iteration, their QUBOs can be processed in parallel, further enhancing computational efficiency.

It is important to note that our estimates of quantum speedup account for the time required to construct the QUBO problem, $T_\text{QUBO}$, i.e., the time needed to generate the QUBO coefficient matrix $\mathbf{Q}$. Because our method produces a sparse $\mathbf{Q}$, $T_\text{QUBO}$ should theoretically scale linearly with problem size. However, in practice, implementation details can influence the actual scalability. Achieving this ideal linear scaling will require an improved code implementation, as discussed in \S\ref{subsec:results-multi}, which we plan to pursue next. With this enhancement, we expect an even greater quantum speedup.

Finally, we have focused on reducing both the total number of iterations and the per-iteration optimization time. The reduction in iteration count effectively decreases the number of repeated PDE solves, thereby lowering the overall computational cost. Further efficiency gains could be realized by accelerating the PDE solving itself. In this regard, integrating quantum-accelerated PDE solvers \cite{harrow2009quantum,QALinearSystem_PRL2019,QA_FEM_CMAME2022} into TO presents a promising avenue for future research, offering the potential to further enhance the efficiency and scalability of TO. \revtwo{Along different lines, integrating with variational quantum algorithms \cite{kim2025variational}, which enable single-loop parallel searches for optimal configurations that satisfy physical constraints through quantum entanglement, offers another potential future direction. In addition, quantum machine learning (QML) approaches could be leveraged to achieve further acceleration, such as the framework proposed by Sukulthanasorn and Terada \cite{SUKULTHANASORN2026118411}. Their approach employs quantum neural networks based on parameterized quantum circuits to learn the complex relationships between the objective function, sensitivity, and design variables, can scale to arbitrary problem sizes with a fixed number of qubits, and has been shown to substantially accelerate optimality criteria (OC)–based TO.}

\section*{CRediT authorship contribution statement}
\textbf{Zisheng Ye}: Writing – original draft, Conceptualization, Methodology, Software, Investigation, Validation, Formal analysis, Data curation. \textbf{Wenxiao Pan}: Writing – original draft, Writing – review \& editing, Supervision, Conceptualization, Methodology, Investigation, Funding acquisition, Resources, Project administration.

\section*{Declaration of competing interest}
The authors declare that they have no known competing financial interests or personal relationships that could have appeared to influence the work reported in this paper.

\section*{Acknowledgment}
This work was supported by the University of Wisconsin - Madison Office of the Vice Chancellor for Research and Graduate Education with funding from the Wisconsin Alumni Research Foundation.

\section*{Data availability}
The codes and examples that support the findings of this study are available in the link: \url{https://github.com/Pan-Group-UW-Madison/QCTO-MDW}.

\revthree{
\appendix
\setcounter{table}{0}
\renewcommand\thetable{\Alph{section}.\arabic{table}}
\section{Prescribed Tolerance in Eq. \eqref{eq:converg_criterion_Lagrange}}\label{sec:sens_tolerance}
In all numerical results presented in this paper, the tolerance in Eq. \eqref{eq:converg_criterion_Lagrange}, which governs the termination of the MDW iterations, was set to $10^{-6}$. To assess the sensitivity of the outer DVTO–MT iteration (the $k$-iteration) to the tolerance prescribed for the inner MDW loop (the $l$-loop), we conducted additional tests with tolerance values ranging from $10^{-3}$ to $10^{-7}$. For illustration, we considered the single-material design by MDW-CC, as described in \S\ref{subsec:results-single}, with the discretization resolution of $100 \times 400 \times 100$. The results are summarized in \Cref{tab:sens_tolerance}. As the tolerance is relaxed, only a slight increase in the total iteration count ($N_{\text{Iter}}$) or the resulting objective function value ($f$) is observed. These results demonstrate that the proposed method is robust with respect to approximations in the Lagrange multipliers. 
\begin{table*}[htbp]
    \centering
    \caption{Sensitivity of the objective function value ($f$) and total iteration count ($N_{\text{Iter}}$) to the tolerance prescribed in Eq. \eqref{eq:converg_criterion_Lagrange}.}    
    \begin{tabular}{ccc}
        \toprule
        Tolerance & $N_{\text{ite}}$ & $f$ \\
        \midrule
        $10^{-7}$ & 41 & 13.571 \\
        $10^{-6}$ & 47 & 13.624 \\
        $10^{-5}$ & 67 & 13.552 \\
        $10^{-3}$ & 53 & 13.948 \\
        \bottomrule
    \end{tabular}
    \label{tab:sens_tolerance}
\end{table*}

\emph{Remarks:} We note that the results reported for the tolerance value $10^{-6}$ differ slightly from those presented in \Cref{tab:single-mat-dw-simp-obj}. This discrepancy is primarily due to that the Gurobi optimizer (v11.0.0) \cite{gurobi} employs stochastic heuristics to accelerate the search for optimal solutions, which prevents exact reproducibility across independent runs. In addition, this set of tests were conducted on a workstation equipped with two Intel\textsuperscript{\textregistered} Xeon\textsuperscript{\textregistered} CPU E5-2698 v4 processors, using 32 MPI processes per run. Variations in domain decomposition associated with the number of processing cores may also introduce minor differences in the results.
}

\bibliographystyle{elsarticle-num}

\biboptions{sort&compress}
\end{document}